\documentclass[useAMS,usegraphicx]{mn2e}

\usepackage{color,soul}
\usepackage{rotating}

\usepackage{bm}
\usepackage{amssymb}
\usepackage{amsmath}
\usepackage{multirow}
\usepackage{twoopt}
\usepackage{times}
\usepackage{color}
\usepackage{units}
\usepackage{subfigure}
\usepackage{version}

\title[Infrared colour properties of nearby radio-luminous galaxies]{Infrared colour properties of nearby radio-luminous galaxies}

\author[]{Xiao-hong Yang$^{1,3}$\thanks{E-mail:yangxh@cqu.edu.cn}, Pei-sheng Chen$^{2}$, and Yan Huang$^{1}$\\
$^{1}$ Department of Physics and Institute of Theoretical Physics, Chongqing University, Chongqing 400044, China\\
$^{2}$ National Astronomical Observatories/Yunnan Observatory, Chinese Academy of Sciences, P.O. Box 110 Kunming, Yunnan Province 650011, China\\
}

\begin{document}

\date{Accepted , Received , in original form }

\pagerange{\pageref{firstpage}--\pageref{lastpage}} \pubyear{2014}

\maketitle

\label{firstpage}

\begin{abstract}
By combining the data of the Two Micron All Sky Survey, the \textit{Wide Field Infrared Survey Explorer} and the \textit{AKARI} satellite, we study the infrared colour properties of a sample of 2712 nearby radio-luminous galaxies (RLGs). These RLGs are divided into radio-loud (RL) active galactic nuclei (AGNs), mainly occurring at redshifts of $0.05<z<0.3$ and star-forming-dominated RLGs (SFGs), mainly occurring at redshifts of $0.01<z<0.15$. RL AGNs and SFGs are separately distributed in the ([3.4]$-$[4.6])$-$([4.6]$-$[12]) two-colour diagram, in which the RL AGNs display a double-core distribution, and the SFGs display a single-core distribution. SFGs have a redder [4.6]$-$[12] colour than RL AGNs due to the significant contribution from the dust component of SFGs. We find simple criteria of MIR colour separation between RL AGNs and SFGs such that: 95$\%$ of RL AGNs have [4.6]$-$[12] $<$ 3.0 and 94$\%$ of SFGs have [4.6]$-$[12] $>$ 3.0. We also analyse the MIR colours of RL AGNs divided into low- and high-excitation radio galaxies (LERGs and HERGs, respectively). The ([3.4]$-$[4.6])$-$([4.6]$-$[12]) diagram clearly shows separate distributions of LERGs and HERGs and a region of overlap, which suggests that LERGs and HERGs have different MIR properties. LERGs are responsible for the double-core distribution of RL AGNs on the ([3.4]$-$[4.6])$-$([4.6]$-$[12]) diagram. In addition, we also suggest 90$-$140$\mu$m band spectral index $\alpha(90,140)<-1.4$ as a criterion of selecting nearby active galaxies with non-thermal emissions at far-infrared wavelengths.
\end{abstract}

\section{Introduction}

Radio-luminous galaxies (RLGs) form a subclass of active galaxies.
The radio emissions from galaxies are thought to originate from star
formation and the accretion process around supermassive black holes
in active galactic nuclei (AGNs). AGNs emit photons over the entire
electromagnetic spectrum and the radio emissions from AGNs are
produced by synchrotron radiation of the relativistic jets and the
lobes formed by interaction of the jets with the surrounding media
(e.g. Antonucci 1984; Urry \& Padovani 1995; Cleary et al. 2007).
The radio emissions from the star-forming region are believed to
originate from free$-$free emission in the ionized hydrogen (H II)
region and the synchrotron radiation by relativistic electrons in
supernova remnants (Harwit \& Pacini 1975). Most active galaxies may
contain two components: AGNs and star-forming regions. For example,
NGC 1068 shows an AGN at the centre surrounded by a ring-shaped
star-forming region (Telesco et al. 1984).

Both AGNs and star-forming regions require a large supply of dusts
and gases as fuel. AGNs have different observed subclasses that can
be explained by a unification model containing a supermassive black
hole surrounded by an accretion disc with jets and a torus of dust
and gas (Antonucci 1993). The dusty torus not only supplies the fuel
of the accretion disc, but also reprocesses X-ray emissions from a
hot corona above the accretion disc and then emits infrared
radiation. Therefore, the infrared bands are important for
understanding the accretion environment of AGNs. In star-forming
regions, dust and gas clouds are heated by young massive stars and
also emit mid-infrared (MIR) and far-infrared (FIR) radiation.

Active galaxies display various properties at various infrared
wavelengths. At near-infrared (NIR) wavelengths,  emission of AGNs
is significant, although the contribution of host galaxies to the
spectral energy distribution (SED) cannot be neglected. For example,
the NIR spectra of quasars are roughly represented by a power-law
(PL) continuum superposed by a composite blackbody (BB) spectrum of
the stellar population (Elvis et al. 1994; Richards et al. 2006;
Park et al. 2010). The composite BB spectrum peaks at $\sim 1.6
\mu$m. Therefore, NIR colour is also used to distinguish AGNs from
normal stars (Kouzuma \& Yamaoka 2010). At MIR wavelengths, dusts in
H II regions (or tori), polycyclic aromatic hydrocarbons (PAHs) in
photodissociation regions and AGNs comprise potential sources of MIR
emission of active galaxies (e.g. Lacy et al. 2004;	Ichikawa et al. 2014). At FIR
wavelengths, it is reported to observe cooler ($\sim 30$ K)
interstellar regions heated by stars in host galaxies or the outer
part of the torus in the AGN (e.g. Ichikawa et al. 2012).

In the present study, we summarize previously reported NIR, MIR and FIR properties of radio galaxies. Strong infrared emission lines have been commonly detected in high-redshift radio galaxies (Rawlings et al. 1989; Eales \& Rawlings 1993; Iwamuro et al. 1996; Evans 1998). For example, a low-resolution NIR spectrum (1.14$-$1.70$\mu$m) of 4C 40.36 ($z=2.269$) shows distinctively strong [O III]$\lambda4959/5007$ doublet emission lines that significantly contribute to the $H$-band image (Iwamuro et al. 1996). A concentric distribution of strong oxygen line-emitting clouds is suggested by the $H$-band image of 4C 40.36 (Iwamuro et al. 1996).

Benefiting from observations implemented by the \textit{Infrared Space Observatory} (\textit{ISO}) and the \textit{Spitzer Space Telescope} , many previous reports have focused on the MIR properties of high-redshift powerful radio galaxies. Siebenmorgen et al. (2004) used ISOCAM data recorded on board the \textit{ISO} to construct the SED of 3CR radio galaxies at redshifts of z$\leq$2.5. They determined that the MIR properties of 3CR radio sources may include synchrotron radiation from AGN, stars of the host galaxy or dust. Cleary et al. (2007) measured the MIR radiation from a sample of extremely powerful 3CRR radio galaxies at redshifts of 0.4$\leq$z$\leq$1.2 based on \textit{Spitzer} observations. They hypothesized that Fanaroff$-$Riley type II quasars are more luminous than galaxies in the MIR range because of a combination of Doppler-boosted synchrotron emissions in quasars and dust extinction in galaxies, both of which are orientation-dependent effects. De Breuck et al. (2010) presented a \textit{Spitzer} survey of 70 radio galaxies at redshifts of 1$<$z$<$5.2 and measured the rest-frame 1.6 $\mu$m emission from the stellar population and the hot dust emission associated with the active nucleus. Leipski et al. (2010) obtained rest-frame 9$-$16 $\mu$m spectra of 11 quasars and 9 radio galaxies from the 3CRR catalogue at redshifts of 1$<$z$<$1.4 based on \textit{Spitzer} observations. They determined that the mean radio-galaxy spectrum shows a silicate absorption feature, whereas the mean quasar spectrum shows a silicate emission feature (Leipski et al. 2010).

Based on \textit{Infrared Astronomical Satellite} (\textit{IRAS}) data, 3CR radio sources were measured at MIR and FIR wavelengths. The results confirm that MIR and FIR emissions from 3CR quasars and radio galaxies are anisotropic and that the non-thermal beamed emission of the core component partly contributes to the FIR emission of the 3CR radio sources (Heckman et al. 1992; Hes et al. 1995). Dicken et al. (2009) presented the deep MIR (24 $\mu$m) to FIR (70 $\mu$m) \textit{Spitzer} photometric observations of a 2Jy sample of powerful radio galaxies (0.05$<$z$<$0.7). They determined that direct AGN heating is energetically feasible for the dust producing both MIR and FIR continua (Dicken et al. 2009). Hardcastle et al. (2010) used \textit{Herschel}-Astrophysical Terahertz Large Area Survey data to analyse the FIR properties of radio galaxies. They reported that these properties are similar to those of a comparison population of radio-quiet galaxies matched in redshift and \textit{K}-band absolute magnitude.

However, little attention has been paid to the infrared colour of
nearby RLGs, the activities of which are dominated by either AGNs or
star formation (Best et al. 2005; Best \& Heckman 2012). This paper
will focus on the infrared colour properties of nearby RLGs,
including that reported by Best et al. (2005) based on sample of
2712 RLGs. The main objective of this study is to examine the
differences in infrared properties of AGN-dominated RLGs and
star-forming region-dominated RLGs by using this large known sample.
Additionally, AGN-dominated RLGs include two different activity
modes (Hardcastle et al. 2007; Best \& Heckman 2012):
high-excitation radio galaxies (HERGs) and low-excitation radio
galaxies (LERGs). HERGs are powered by radiatively efficient
standard thin discs at high accretion rates, whereas LERGs are
fuelled at low accretion rates by radiatively inefficient accretion
flows, i.e. advection-dominated accretion flows (ADAFs; Narayan \&
Yi 1995; Hardcastle et al. 2007; Best \& Heckman 2012; Yuan \& Narayan 2014). G\"{u}rkan
et al. (2014) analysed the MIR properties of HERGs and LERGs by
using four complete samples (3CRR, 2Jy, 6CE and 7CE). They pointed
out that HERGs are separated from LERGs in the MIR-radio plane
(G\"{u}rkan et al. 2014). We further analyse the MIR colours of
HERGs and LERGs by using a sample of nearby RLGs.

Throughout this paper, we adopt a flat universe with a Hubble constant $H_0=70 \text{ km s}^{-1} \text{Mpc}^{-1}$, $\Omega_{\Lambda}=0.7$ and $\Omega_{\text{m}}=0.3$.

The paper is organized as follows: Section 2 describes our data; Sections 3$-$5 discuss the colour
properties; and Section 6 provides a summary.


\begin{table*}
\begin{center}

\caption[]{Infrared photometric data of nearby radio-luminous galaxies (RLGs)} \label{table1} \tiny
\begin{tabular}{ccccccccccccc}
\hline\noalign{\smallskip} \hline\noalign{\smallskip}
RA(2000)\text{      }Dec.(2000) & $z$ & $E$($B-V$) & $J$ (mag) &
 $H$ (mag)  &
 $K_s$ (mag)  &
 $W1$ (mag)  &
 $W2$ (mag)  &
 $W3$ (mag)  &
 $W4$ (mag) &
 $90\mu m$ (Jy) &
 $140 \mu m$ (Jy) \\
 (1)  & (2)  &(3) &
 (4)  &  (5)  &
 (6)  &  (7)  &
 (8)  &  (9)  &
 (10) &  (11) &
 (12) \\
 \hline\noalign{\smallskip}
146.9561  -0.3423    &     0.1347  & 0.1343 &  15.882(0.109) & 15.120(0.160) & 14.526(0.119) & 14.054(0.035) & 13.829(0.044) & 12.040(0.000) &   8.452(0.000) &         /           &            /          \\
146.1436  -0.7416    &     0.2039  & 0.0606 &  15.654(0.115) & 14.982(0.133) & 14.416(0.135) & 13.477(0.025) & 13.217(0.030) & 12.156(0.000) &   9.056(0.000) &         /           &            /          \\
146.7371  -0.2522    &     0.1305  & 0.1327 &  15.482(0.094) & 14.798(0.104) & 14.209(0.093) & 13.505(0.026) & 13.320(0.032) & 12.031(0.274) &   8.996(0.000) &         /           &            /          \\
146.3738  -0.3684    &     0.0529  & 0.1118 &  15.137(0.094) & 14.474(0.130) & 13.948(0.087) & 12.655(0.024) & 12.492(0.025) &  8.622(0.029) &   6.205(0.068) &         /           &            /          \\
145.6012  -0.0014    &     0.1459  & 0.0625 &  15.958(0.115) & 14.970(0.099) & 14.158(0.079) & 13.337(0.024) & 12.710(0.025) &  8.381(0.024) &   6.086(0.053) & 0.5394E+0(0.572E-1) &            /          \\
146.4630   0.6387    &     0.0303  & 0.1962 &  15.412(0.143) & 14.725(0.200) & 14.232(0.149) & 12.511(0.023) & 12.313(0.024) &  8.115(0.022) &   5.770(0.044) & 0.4890E+0(0.127E+0) &            /          \\
146.8068   0.6656    &     0.0201  & 0.1484 &  13.764(0.099) & 12.936(0.125) & 12.150(0.064) & 10.861(0.022) & 10.639(0.020) &  6.588(0.014) &   4.592(0.022) & 0.2862E+1(0.192E+0) &  0.4211E+1(0.174E+0)  \\
146.7991   0.7027    &     0.0305  & 0.1495 &  15.785(0.154) & 15.201(0.177) & 14.700(0.167) & 13.204(0.024) & 12.950(0.028) &  9.005(0.033) &   6.099(0.058) & 0.5410E+0(0.374E-1) &  0.2857E+1(0.519E+0)  \\
146.7815   0.7380    &     0.2618  & 0.1485 &  16.337(0.172) & 15.473(0.174) & 14.664(0.144) & 14.134(0.042) & 13.850(0.054) & 11.885(0.000) &   8.268(0.000) &         /           &            /          \\
147.0805   0.7880    &     0.2111  & 0.1039 &  15.820(0.097) & 14.956(0.109) & 14.284(0.097) & 13.734(0.026) & 13.446(0.035) & 11.844(0.000) &   8.758(0.000) &         /           &            /          \\
149.1699  -0.0233    &     0.1392  & 0.0299 &  15.396(0.090) & 14.640(0.107) & 14.175(0.086) & 13.305(0.025) & 13.108(0.029) & 11.888(0.268) &   8.592(0.000) &         /           &            /          \\
148.4325  -1.0264    &     0.1103  & 0.0426 &  15.413(0.086) & 14.569(0.107) & 13.851(0.074) & 12.767(0.023) & 11.460(0.022) &  8.173(0.022) &   5.372(0.036) & 0.5702E+0(0.239E-1) &            /          \\
148.2377  -0.7920    &     0.0898  & 0.0334 &  14.466(0.073) & 13.847(0.094) & 13.373(0.077) & 12.570(0.023) & 12.507(0.025) & 11.690(0.219) &   8.416(0.000) &         /           &            /          \\
147.7086  -0.8878    &     0.2715  & 0.0549 &  16.059(0.113) & 15.369(0.145) & 14.596(0.138) & 13.881(0.026) & 13.544(0.032) & 12.440(0.490) &   8.950(0.000) &         /           &            /          \\
147.4283  -0.8401    &     0.0809  & 0.0651 &  15.132(0.080) & 14.426(0.082) & 13.978(0.076) & 13.298(0.025) & 13.016(0.028) & 11.140(0.131) &   8.647(0.000) &         /           &            /          \\
\hline\noalign{\smallskip}
\end{tabular}

\begin{list}{}
\item\scriptsize{Col. 1: coordinates in the epoch of 2000 from Best et al. (2005); Col. 2: redshifts of sources; Col. 3: derived $E$($B-V$); Cols. 4$-$10: magnitudes at $J$, $H$, $K_s$, $W$1, $W$2, $W$3 and $W$4 bands, respectively, with their uncertainty given by the values in parentheses; Cols. 11 and 12: fluxes in units of Jansky at 90 and 140$\mu$m bands, respectively, with their uncertainty given by the values in parentheses. When the value in parentheses is zero, the listed magnitudes are the upper limits. We list only the sources in which 90- and 140-$\mu$m fluxes have high quality such that the source is confirmed and flux is reliable. Only the first 15 sources are listed here; the full table is available in the electronic version.}
\end{list}

\end{center}
\end{table*}

\begin{figure}

\scalebox{0.5}[0.5]{\rotatebox{0}{\includegraphics[bb=50 40 531
436]{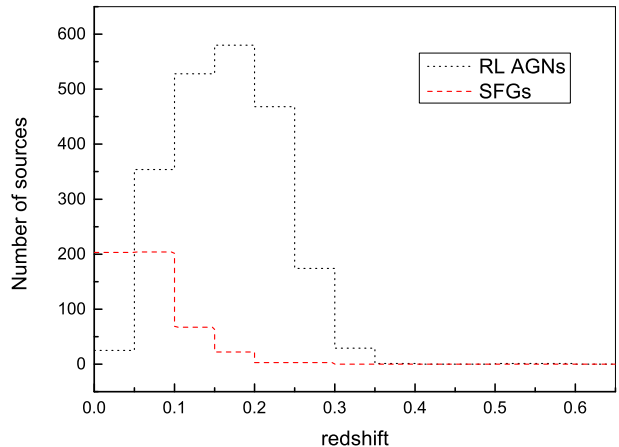}}}

\ \centering \caption{Redshift distribution of all sources studied.
Black line represents radio-loud AGNs (RL AGNs); red line represents star-forming galaxies (SFGs).}
(A color version of this figure is available in the online journal.)
\label{fig 1}
\end{figure}

\section{Data}
\subsection{Sample}

All sources studied here were selected from a catalogue of 2712 RLGs (Best et al. 2005). This large sample of RLGs was derived by Best et al. (2005), who used the second data release of the Sloan Digital Sky Survey (York et al. 2000; Stoughton et al. 2002) to cross-correlate with combined data of the National Radio Astronomy Observatories Very Large Array Sky Survey (Condon et al. 1998) and the Faint Images of the Radio Sky at 20 cm (Becker et al. 1995) survey. Both radio surveys achieved higher angular resolutions than those recorded by their predecessors and covered substantial fractions of the sky, down to flux densities at the millijansky scale. To remove the very extended galaxies, Best et al. (2005) excluded galaxies with redshifts of $z < 0.01$; the 2712 RLGs are located at redshifts of $0.01<z<0.56$. These samples are from the SDSS main galaxy catalogue, which means that objects classified as ¡®quasars¡¯ by the automated SDSS classification pipeline are excluded (Best et al. 2005).

\begin{table}
\begin{center}

\caption[]{Number of detections and upper limits for low-excitation radio galaxies (LERGs), high-excitation radio galaxies (HERGs) and star-forming galaxies (SFGs).} \label{table3}\tiny
\begin{tabular}{ccccccccc}

\hline\noalign{\smallskip}
 Subsample       & $W$1(d) & $W$1(u) & $W$2(d) & $W$2(u) & $W$3(d) & $W$3(u)& $W$4(d) & $W$4(u) \\
\hline\noalign{\smallskip}
LERGs & 1745 &3  & 1748 &0  & 1032 &716 & 220 &1528 \\
HERGs & 52   &0  & 52   &0  & 45   & 7  & 32  &20 \\
SFGs  & 502  &0  & 502  &0  & 501  & 1  & 490 &12 \\

\hline\noalign{\smallskip}
\end{tabular}

\begin{list}{}
\item\scriptsize{Cols 2$-$3, 4$-$5, 6$-$7 and 8$-$9 give the number of detections and upper limits for each class at four bands of $W$1, $W$2, $W$3 and $W$4, respectively. The `d' in parenthesis indicates the number of detections, whereas the `u' indicates the number of upper limits. }
\end{list}

\end{center}
\end{table}

\subsection{Cross-identification}

In recent years, many infrared surveys have been completed by various instruments at various infrared bands. The Two Micron All Sky Survey (2MASS) has swept 99.998$\%$ of the whole sky at the $J$ (1.25$\mu$m), $H$ (1.65$\mu$m) and $K_s$ (2.16$\mu$m) bands (Skrutskie et al. 2006). The \textit{Wide-Field Infrared Survey Explorer} (WISE) has completed an all-sky survey at the $W$1 (3.4$\mu$m), $W$2 (4.6$\mu$m), $W$3 (12$\mu$m) and $W$4 (22$\mu$m) bands and issued two public data releases in April 2011 and March 2012. The WISE achieves 5$\sigma$ point-source sensitivities better than 0.08, 0.11, 1 and 6 mJy in the four bands, corresponding to Vega magnitudes of 16.5, 15.5, 11.2 and 7.9, respectively; the sensitivity of the WISE $W$3 (12$\mu$m) band is more than 100 times better than that of \textit{IRAS} (Wright et al. 2010). The \textit{AKARI} (Ishihara et al. 2010), a Japanese infrared space mission, has carried MIR to FIR surveys at six bands centred at 9, 18, 65, 90, 140 and 160 $\mu$m by using two instruments: an infrared camera (Onaka et al. 2007) and a FIR surveyor (FIS; Kawada et al. 2007). The \textit{AKARI} achieves higher sensitivity, higher spatial resolution and wider wavelength coverage than the \textit{IRAS}.

To obtain the NIR, MIR and FIR photometric data of the studied radio sources, we primarily used the point source catalogues (PSCs) of 2MASS, WISE and \textit{AKARI} (Skrutskie et al. 2006; Ishihara et al. 2010; Yamamura et al. 2010). The catalogue of the 2712 RLGs is cross-identified with the 2MASS, WISE and \textit{AKARI}-FIS PSCs. Most 2MASS PSC sources have spatial resolutions of less than $\sim$4 arcsec in all of the three wavelength bands (see 2MASS homepage\footnote{http://www.ipac.caltech.edu/2mass/}). The spatial resolutions of WISE are 6.1, 6.4, 6.5 and 12.0 arcsec at $W$1, $W$2, $W$3 and $W$4, respectively (Wright et al. 2010). The spatial resolutions of \textit{AKARI} are 7 and 48 arcsec for the MIR and FIR instruments, respectively (e.g. Matsuta et al. 2012). We used position offsets of 4, 6 and 20 arcsec to search the 2MASS, WISE and AKARI-FIS PSC counterparts, respectively, for the 2712 RLGs. If more than one object was within the position offset radius, the closest object was accepted as the counterpart. In practice, only a small number of RLGs fell into the category. 96$\%$ of the 2MASS PSC counterparts and 90$\%$ of the WISE PSC counterparts were within a 1 arcsec radius of the RLGs; 91$\%$ of the \textit{AKARI}-FIS counterparts were within a 10 arcsec radius.

Finally, we excluded RLGs without 2MASS counterparts and obtained a sample of 2663 RLGs. About 42$\%$ of the studied samples were consistent with a point source. Figure 1 shows the redshift distribution of two types of RLGs: radio-loud (RL) AGNs and star-forming galaxies (SFGs). The classification of RLGs is described in section 2.4. The RL AGNs studied are located mainly at redshifts of  $0.05<z<0.3$, whereas the SFGs studied are located mainly at redshifts of $0.01<z<0.15$. The infrared photometric data of the 2663 RLGs are listed in Table 1, in which Col. 1 gives the coordinates (in the epoch of 2000) from Best et al. (2005); Col. 2 gives the redshifts of the sources; Col. 3 gives the derived E(B$-$V); Cols. 4$-$10 give the vega magnitudes at the $J$, $H$, $K_s$, $W$1, $W$2, $W$3 and $W$4 bands, respectively, with their uncertainties given by the values in parentheses (Skrutskie et al. 2006; Ishihara et al. 2010); Cols. 11 and 12 give the flux values in units of Janskies at the 90 and 140$\mu$m bands, respectively, with their uncertainties given by the values in parentheses (Yamamura et al. 2010). Only the first 15 sources are listed here; the full table is available in the electronic version of this paper. Table 2 also gives number of detections and upper limits for LERGs, HERGs and SFGs.

\subsection{Galactic interstellar extinction correction}
The interstellar extinction corrections caused by our Galaxy, $E$($B-V$), are considered in five bands: $J$, $H$, $K_s$, $W$1 and $W$2. We obtain the values of $E$($B-V$) for all samples through the NASA/IPAC extragalactic database (NED Home Page\footnote{http://irsa.ipac.caltech.edu/applications/DUST/}), which gives the Galactic dust reddening for a line of sight based on Schlegel, Finkbeiner and Davis (1998). The extinction of a given wavelength band $a$ is usually estimated by $A(a)=R(a)\times E(B-V)$. Yuan et al. (2013) reported new empirical extinction coefficients $R(a)$. According to Table 2 of Yuan et al. (2013), we employed $R(a)=0.72$, $0.46$, $0.306$, $0.19$, and $0.15$ for the bands $J$, $H$, $K_s$, $W$1 and $W$2, respectively, to correct the magnitudes of $J$, $H$, $K_s$, $W$1 and $W$2. In fact, the extinctions were smaller than the uncertainties of the measured magnitudes for the majority of the sources. Therefore, we applied interstellar extinction corrections for only about 2 \% of our samples.

\subsection{Classification}

At NIR wavelengths, thermal emissions from host galaxies and non-thermal emissions from the central nuclei were observed. The orientation of AGNs may affect non-thermal emissions from the central nuclei. Therefore, to investigate the effect of orientation, we used the 13th version of Quasars and AGNs (V\'{e}ron-Cetty \& V\'{e}ron 2010) to search the spectral types of our sample and obtain the sources sub-classified as Seyfert 1 (Sy1) and Setfert 2 (Sy2) galaxies.

At MIR and FIR wavelengths, thermal emissions from dust play a very important role in RLGs. To further investigate the MIR properties of different active types of RLGs, we also distinguished the active types of RLGs according to Best et al. (2005) and Best \& Heckman (2012). Based on the location of a galaxy in a plane of 4000-{\AA} break strength ($D_{n}(4000)$) versus radio luminosity per unit stellar mass ($L_{1.4GHz}/M_{*}$), the RLGs were divided into RL AGNs and SFGs (Best et al. 2005). The radio emissions of RL AGNs are mainly generated by jets or lobes, whereas those of SFGs are dominated by star formation. The majority of the samples used by Best et al. (2005) are also included in Best \& Heckman (2012), although a small number of samples were inconsistently classified. For the samples overlapping both works, we employed the classification from Best \& Heckman (2012). Best \& Heckman (2012) further divided RL AGNs into HERGs and LERGs; we employed this classification as well. HERGs and LERGs are believed to belong to different accretion modes, i.e. standard thin discs and ADAFs, respectively (Hardcastle, Evans \& Croston 2007). HERGs have higher accretion rates than LERGs (Best \& Heckman 2012).

\section{NIR properties}
Figure 2 shows a $(J-H)-(H-K_s)$ two-colour diagram (top panel) and the normalized distributions of the colours  $H-K_s$ (middle panel) and $J-H$ (bottom panel). In order to easily understand the spectral shapes of the RLGs, the colours were also transformed to their corresponding spectral indices $\alpha(\lambda_1,\lambda_2)$. Following Sekiguchi (1987), the spectral index is defined as,
\begin{equation}
\alpha(\lambda_1,\lambda_2)=\text{log}(F_{\lambda_1}/F_{\lambda_2})/\text{log}(\lambda_2/\lambda_1),
\end{equation}
where $F_{\lambda_1}$ and $F_{\lambda_2}$ are the measured flux densities at wavelengths $\lambda_1$ and $\lambda_2$ ($\lambda_2>\lambda_1$), respectively. It is noted that the flux densities are expressed throughout this paper by using frequency units. The corresponding spectral indices of the colours $J-H$ and $H-K_s$ are also marked in the top panel of Figure 2. In the two-colour diagram, a PL ( $F_{\lambda}\propto \lambda^{-\alpha}$) line (dashed line) and a BB line (dotted line) are also shown. The PL line can be obtained by calculating the change in colour with the spectral index ($\alpha$). For a PL spectrum, the colour in a certain wavelength range $\Delta \lambda=\lambda_2-\lambda_1$ is derived by
\begin{equation}
m_{\lambda_1}-m_{\lambda_2} = -2.5 \text{ log}[(\frac{\lambda_1}{\lambda_2})^{-\alpha}(\frac{F_{0 \lambda_2}}{F_{0 \lambda_1}})],
\end{equation}
where $m_{\lambda_1}$ and $m_{\lambda_2}$ are magnitudes at wavelengths $\lambda_1$ and $\lambda_2$, respectively, and $F_{0 \lambda_1}$ and $F_{0 \lambda_2}$ are absolute flux calibrations at wavelengths $\lambda_1$ and $\lambda_2$, respectively.  Similarly, we calculate the change in colour with temperature ($T$) to draw the BB line. For a BB spectrum, the colour is derived by
\begin{equation}
m_{\lambda_1}-m_{\lambda_2} = -2.5 \text{ log}[(\frac{B_{\lambda_1}}{B_{\lambda_2}})(\frac{F_{0 \lambda_2}}{F_{0 \lambda_1}})],
\end{equation}
where $B_{\lambda_1}$ and $B_{\lambda_2}$ are the Planck function with same temperature at wavelengths $\lambda_1$ and $\lambda_2$, respectively.

\begin{figure}

\scalebox{0.5}[0.5]{\rotatebox{0}{\includegraphics[bb=50 40 531
436]{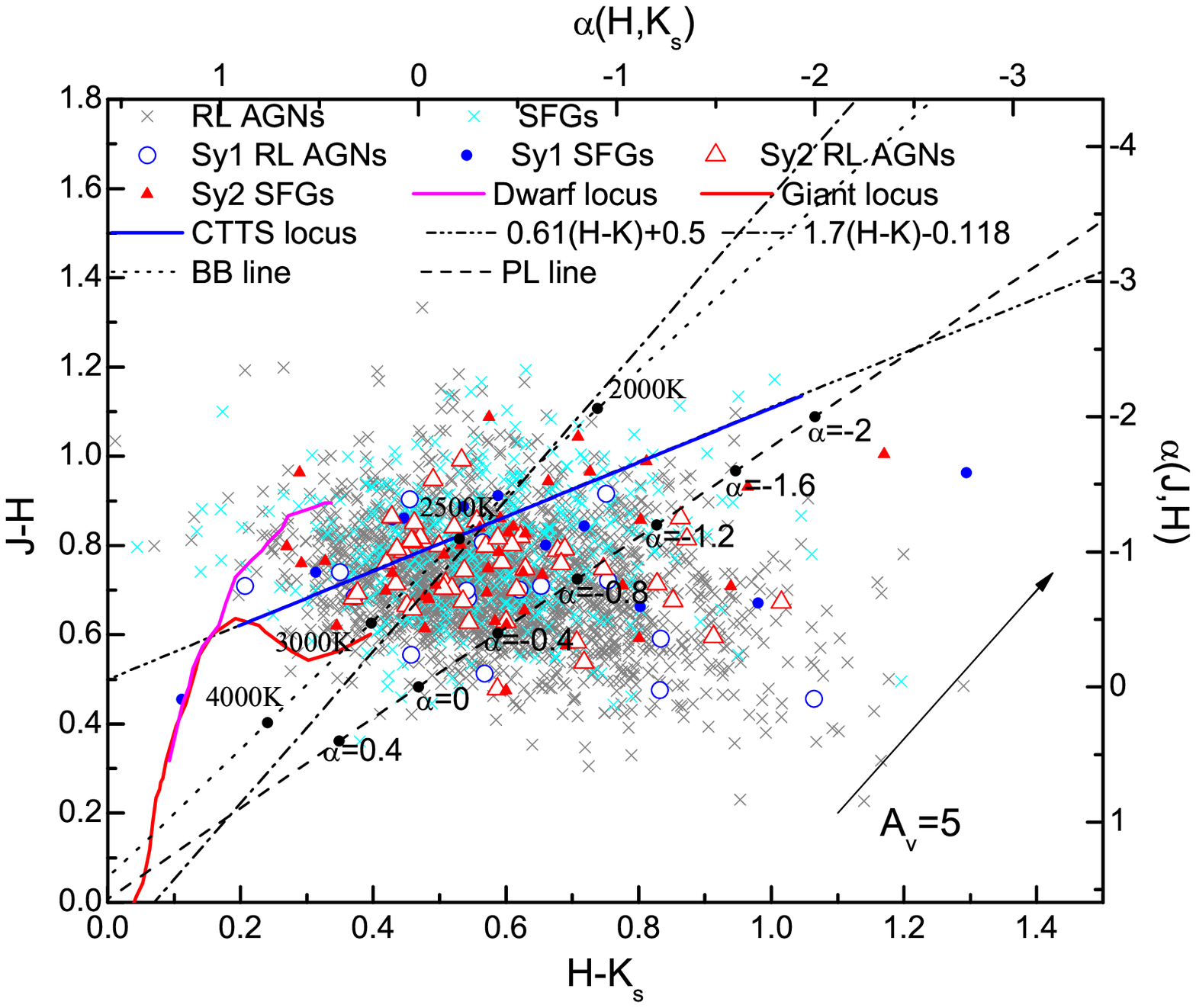}}}

\scalebox{0.5}[0.5]{\rotatebox{0}{\includegraphics[bb=50 40 531
410]{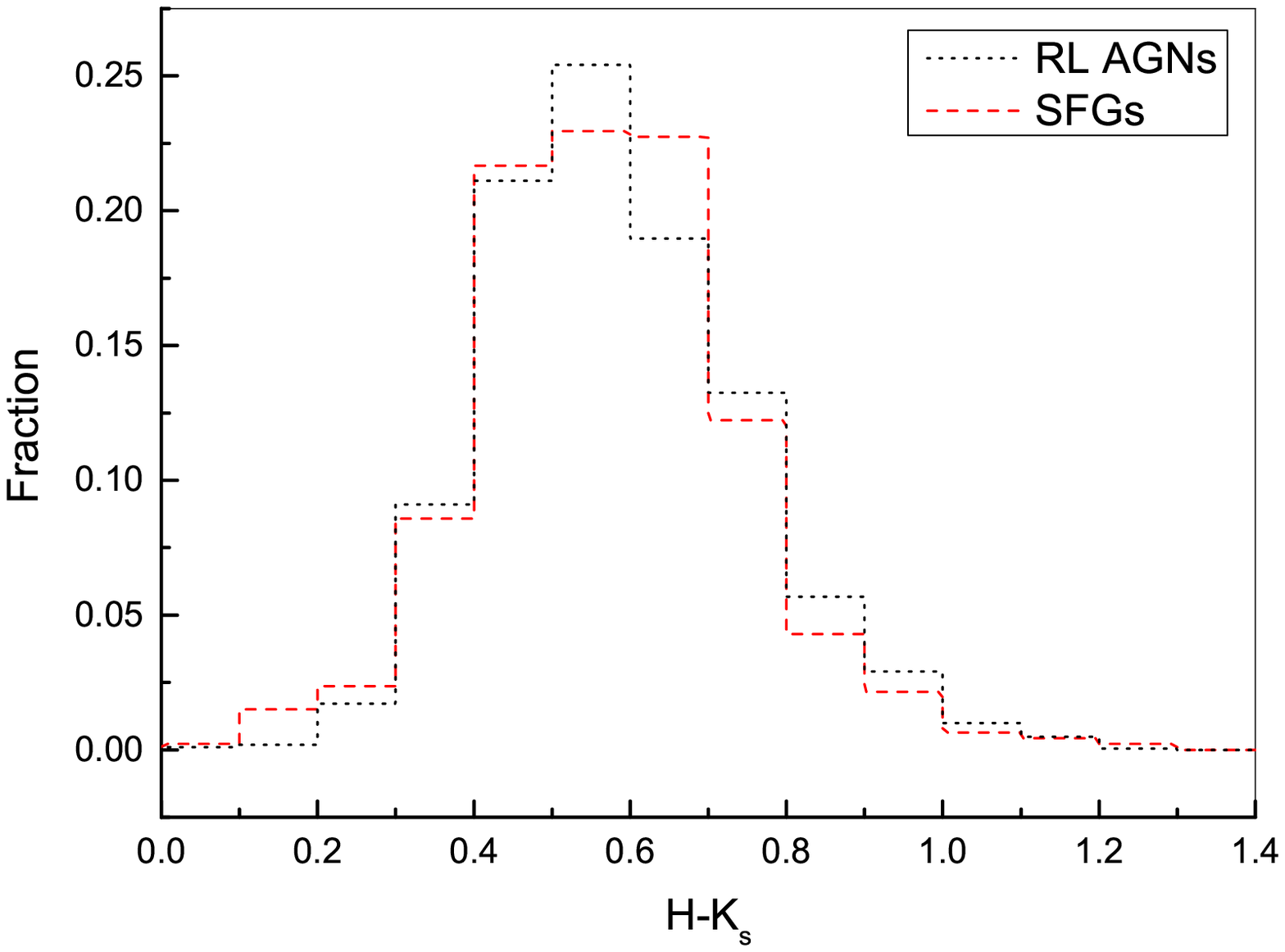}}}
\scalebox{0.5}[0.5]{\rotatebox{0}{\includegraphics[bb=50 40 531
410]{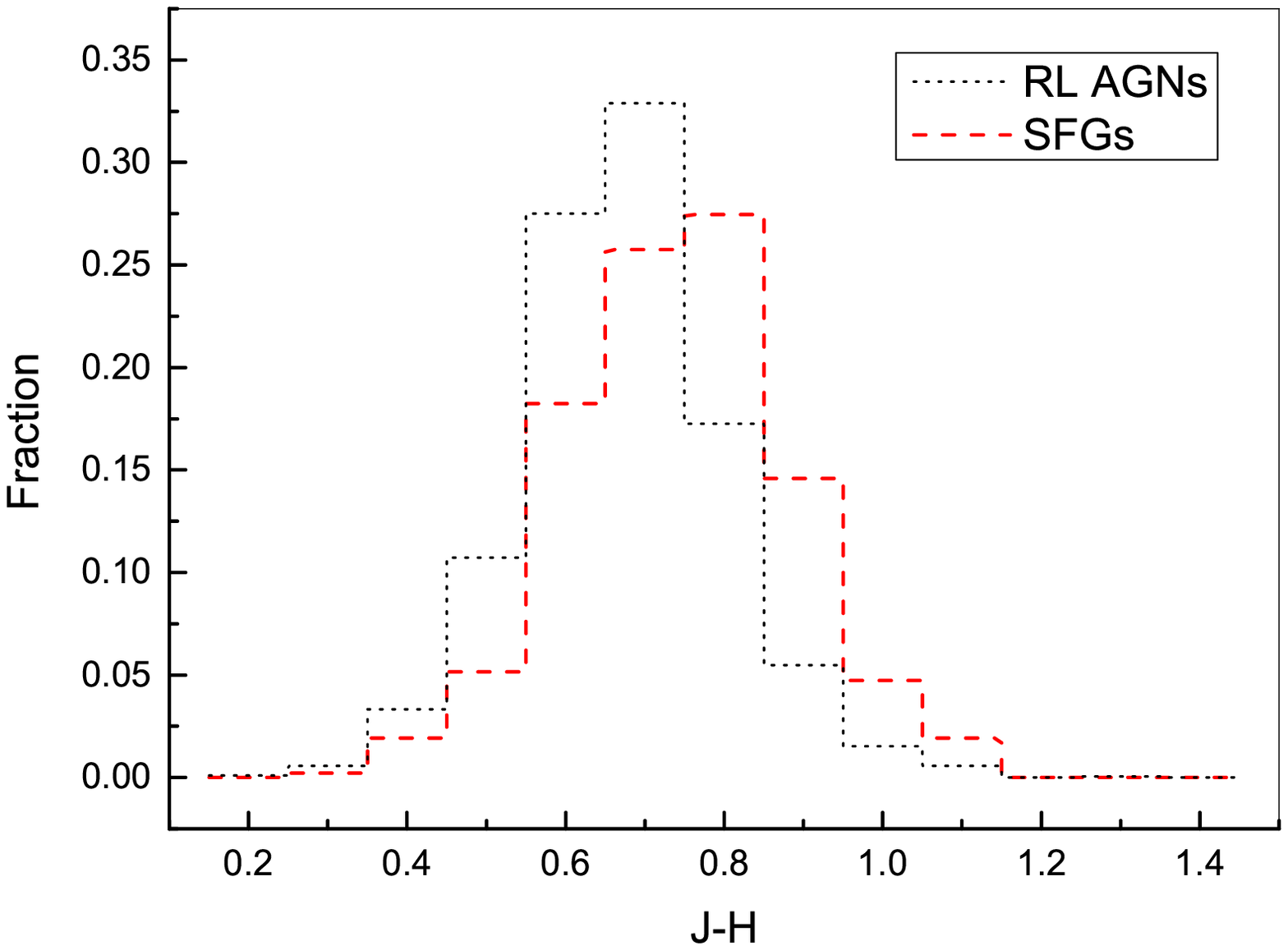}}}

\ \centering \caption{($J-H$)$-$($H-K_s$) two-colour diagram (top panel) and the normalized distribution of the colour $H-K_s$ (middle panel) and J-H (bottom panel). The power-law spectral indices $\alpha(J,H)$ and $\alpha(H,K_s)$ are also marked in the top panel. A BB line with temperatures and a PL line with spectral indices are plotted as the dotted and dashed line, respectively. In the middle and bottom panels, black and red lines represent RL AGNs and SFGs, respectively. }
\label{fig 2}
\end{figure}

In addition, following Kouzuma \& Yamaoka (2010), the top panel of Figure 2 shows the intrinsic loci of dwarf stars (magenta line), giant stars (red line) and Classical T Tauri Stars (CTTS; blue line) according to Bessell \& Brett (1988) and Meyer et al. (1997). The reddening vector from Rieke \& Lebofsky (1985), which indicates the shifting direction of the stellar and CTTS loci on the $(J-H)-(H-K_s)$ diagram, is also plotted. The infrared colour of the RLGs is different from the colours of various types of stars. Kouzuma \& Yamaoka (2010) proposed the following NIR colour criteria for selecting AGNs:
\begin{equation}
(J-H)\leq 1.7 (H-K_s)-0.118,
\end{equation}
\begin{equation}
(J-H)\leq 0.61 (H-K_s) + 0.5.
\end{equation}
Equations 4 and 5 are also indicated in the two-colour diagram by dot$-$dashed and double-dot-dashed lines, respectively. 64$\%$ of RL AGNs and 48$\%$ of SFGs achieved these selection criteria, which is a lower percentage than that given by Kouzuma \& Yamaoka (2010) for low-redshift AGNs. It is noted that the sample used in the present study was located at redshifts of 0.01$<z<$0.56 where the NIR radiation from the centre of the AGN should be contaminated by host galaxies with intrinsic colours similar to those of normal stars. On the contrary, the majority of the samples studied by Kouzuma \& Yamaoka were quasars that had more luminous cores than those of our sample. Hence, the NIR selection was less effective for the RL AGNs than for other types of AGNs.

It should be noted that most of the sources achieving the selection criteria were distributed in the region around the PL line, which indicates that their NIR spectra can be approximately described by a PL spectrum. The top panel of Figure 2 also shows that the majority of the RL AGNs and SFGs had indices of $-2<\alpha(H,K_s)<1$ and
$-2<\alpha(J,H)<0.4$, which is less than the value of 4 supposed by the Rayleigh$-$Jeans spectra. AGNs produced emissions at nearly all wavelengths, and their emissions were dominated by non-thermal radiation. The ultraviolet to MIR continuum of AGNs is represented by a PL (Stern et al. 2005). However, host galaxies may contaminate the NIR colour of AGNs. The spectra from the star population in the host galaxy were determined by using the BB spectra and peak at approximately 1.6 $\mu$m (Elvis et al. 1994; Stern et al. 2005; Richards et al. 2006; Park et al. 2010). For example, Dultzin-Hacyan \& Benitz (1994) pointed out that the dominant NIR emissions of Sy2 galaxies are primarily produced by evolved massive red giants and supergiants in the host galaxies; De Breuck et al. (2010) used the spectrum of a stellar population as a component in fitting the NIR and MIR spectra of high-redshift radio galaxies. To study the degree in which objects with non-thermal emissions and stellar populations affect the NIR colour of active galaxies, Kouzuma \& Yamaoka (2010) first examined the distribution of four types of objects in the $(J-H)-(H-K_s)$ diagram including microquasar candidates, low-mass X-ray binaries, cataclysmic variables and massive young stellar objects and found that most of the objects were not close to the PL line and did not achieve the NIR colour criteria of AGNs. Based on these results, they determined that the dominant NIR radiation of the four objects was likely thermal emission. They further examined the loci of normal galaxies without considering the contribution from the nuclei in the $(J-H)-(H-K_s)$ diagram and concluded that the normal galaxies with $0.6 <\textit{z}< 0.8$ have intrinsic colours similar to those of normal stars, whereas galaxies with $\textit{z}\sim 1$ achieve the NIR colour criteria of AGNs. Therefore, the probable objects with non-thermal radiation and the stellar populations in the host galaxies cannot cause active galaxies with $\textit{z}\lesssim 0.8$ to approach the PL line. Thus, if active galaxies with $\textit{z}\lesssim 0.8$ are close to the PL line in the $(J-H)-(H-K_s)$  diagram, their NIR emissions should be dominated by non-thermal emissions associated with the nuclei. For the SFGs achieving the selection criteria (48$\%$ of SFGs), the non-thermal emissions from the nuclei dominate their NIR emissions. It should also be noted that a great number of sources are distributed in the region around the BB line, which indicates that their NIR emissions may be dominated by thermal radiation because the BB line is considered as an indicator of thermal radiation from the starlight in the host galaxy. The contribution ratio of the centre nuclei and their host galaxies to NIR emissions determines the position of active galaxies in the $(J-H)-(H-K_s)$ diagram.

\begin{figure}

\scalebox{0.5}[0.5]{\rotatebox{0}{\includegraphics[bb=50 40 531
436]{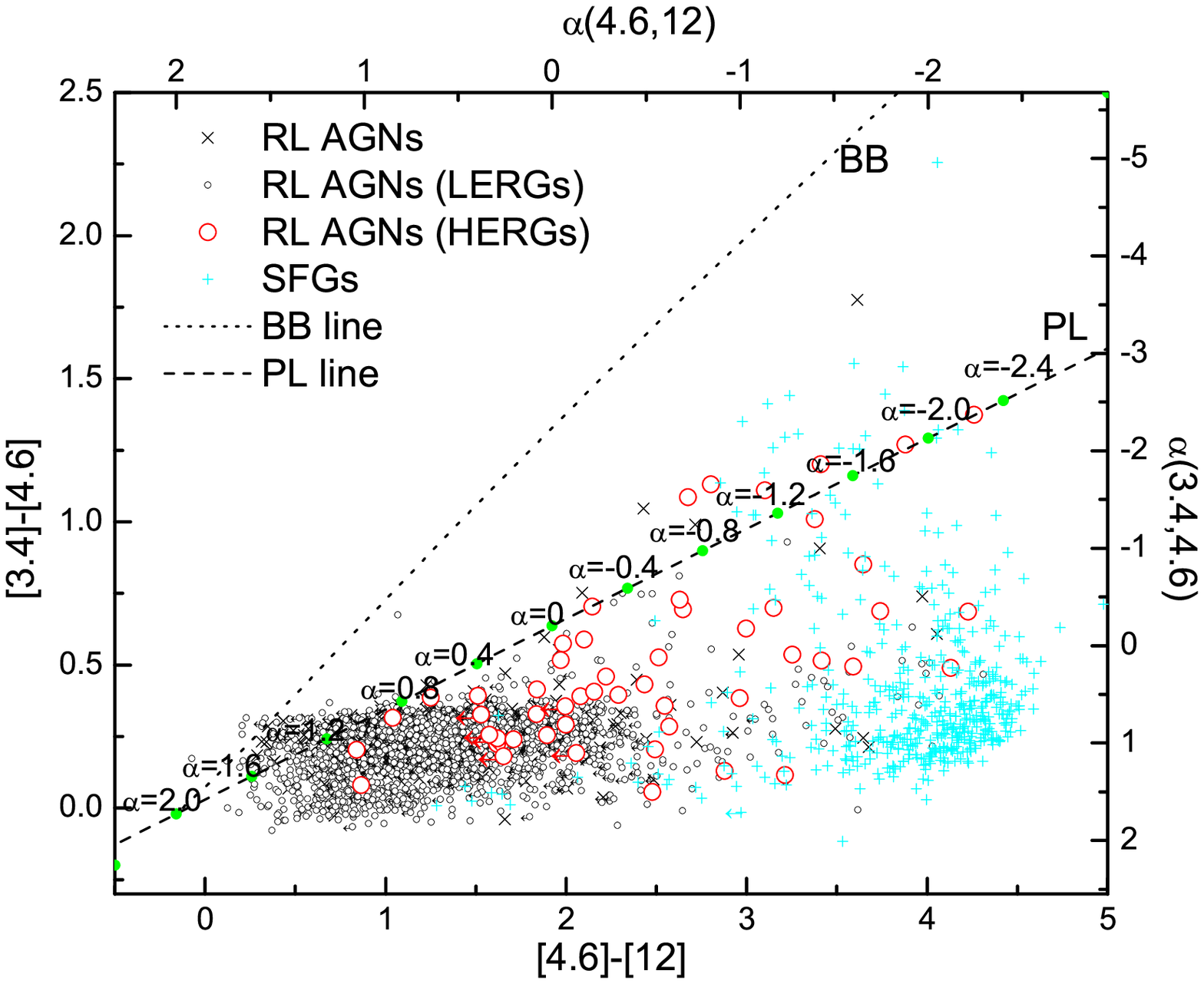}}}

\scalebox{0.5}[0.5]{\rotatebox{0}{\includegraphics[bb=50 40 531
436]{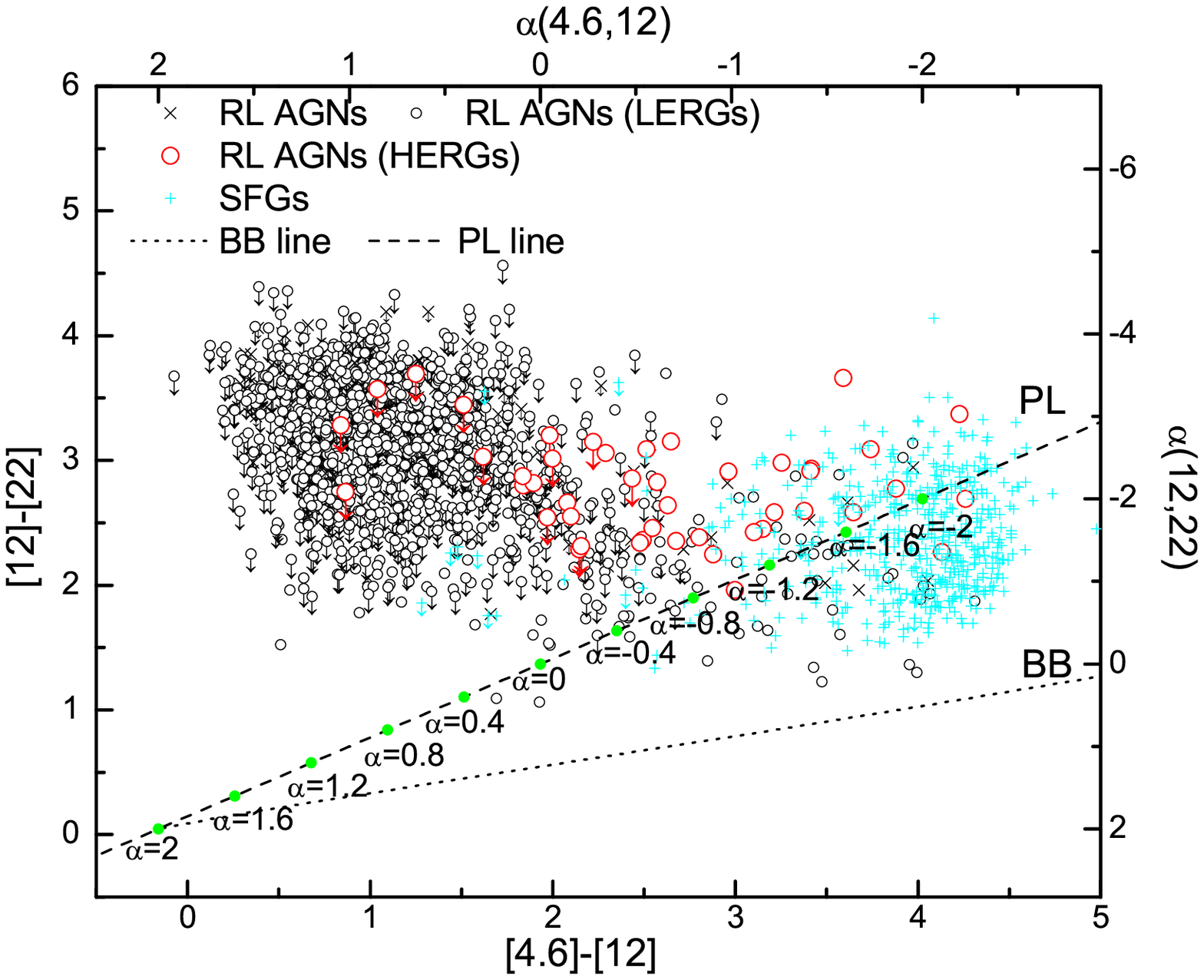}}}

\ \centering \caption{([3.4]$-$[4.6])$-$([4.6]$-$[12]) (top panel) and ([12]$-$[22])-([4.6]$-$[12]) (bottom panel) two-colour diagrams. A PL line marked with spectral indices and a BB line are also plotted by dashed and dotted lines, respectively. The upper limits are indicated as arrows.}
\label{fig 3}
\end{figure}

\begin{figure}

\scalebox{0.5}[0.5]{\rotatebox{0}{\includegraphics[bb=50 40 531
436]{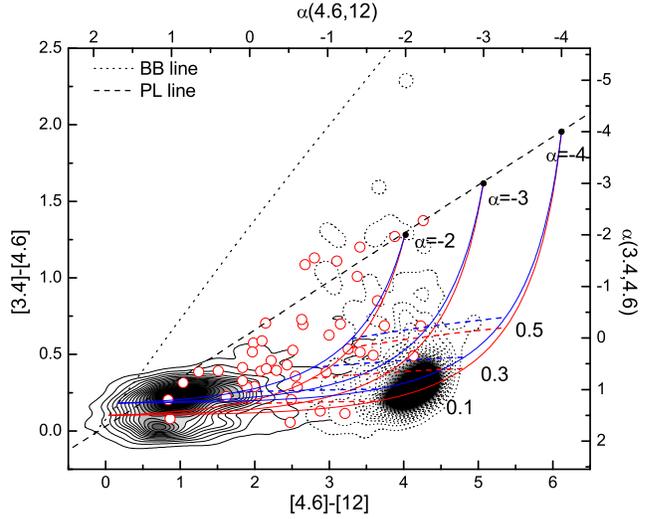}}}

\ \centering \caption{Probability distribution of LERGs and SFGs. Grey-scale and 30 isodensity contour lines (black solid lines: LERGs; black dotted lines: SFGs) indicate the probability distribution of LERGs and SFGs. HERGs are superimposed as red circles in this figure for comparison. A PL line marked with different spectral indices and a BB line are also plotted with dashed and dotted lines, respectively. Blue solid lines indicate the loci of models with $T=3500 K$ and $\alpha = -2$, $-3$ and $-4$. Red solid lines indicate the loci of models with $T=5500 K$ and $\alpha = -2$, $-3$ and $-4$. Numbers along the loci are the fraction of contribution of the PL component to the $W$2-band flux. Objects with upper limits are excluded.}
\label{fig 4}
\end{figure}

\begin{figure}

\scalebox{0.5}[0.5]{\rotatebox{0}{\includegraphics[bb=50 40 531
436]{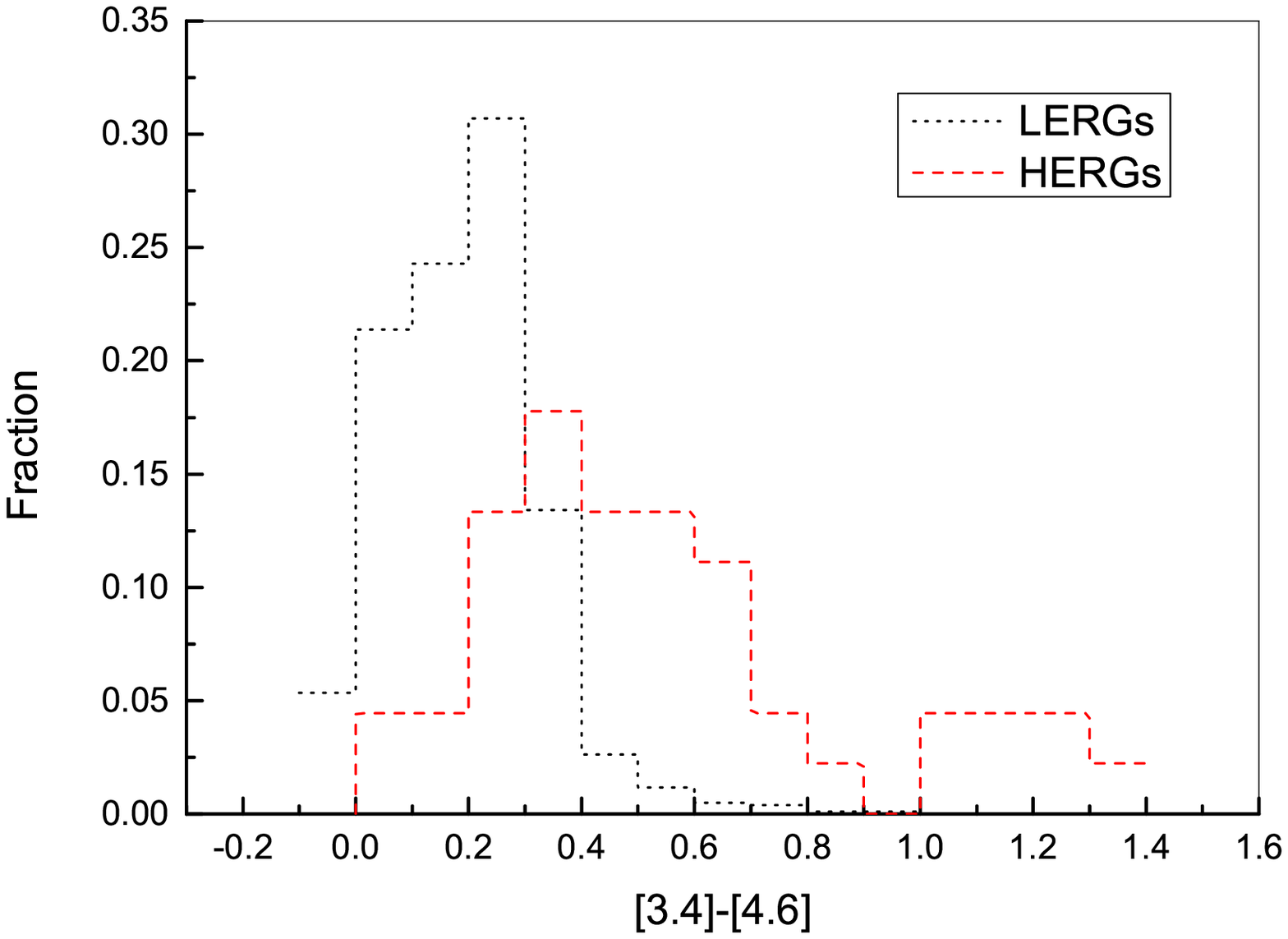}}}
\scalebox{0.5}[0.5]{\rotatebox{0}{\includegraphics[bb=50 40 531
436]{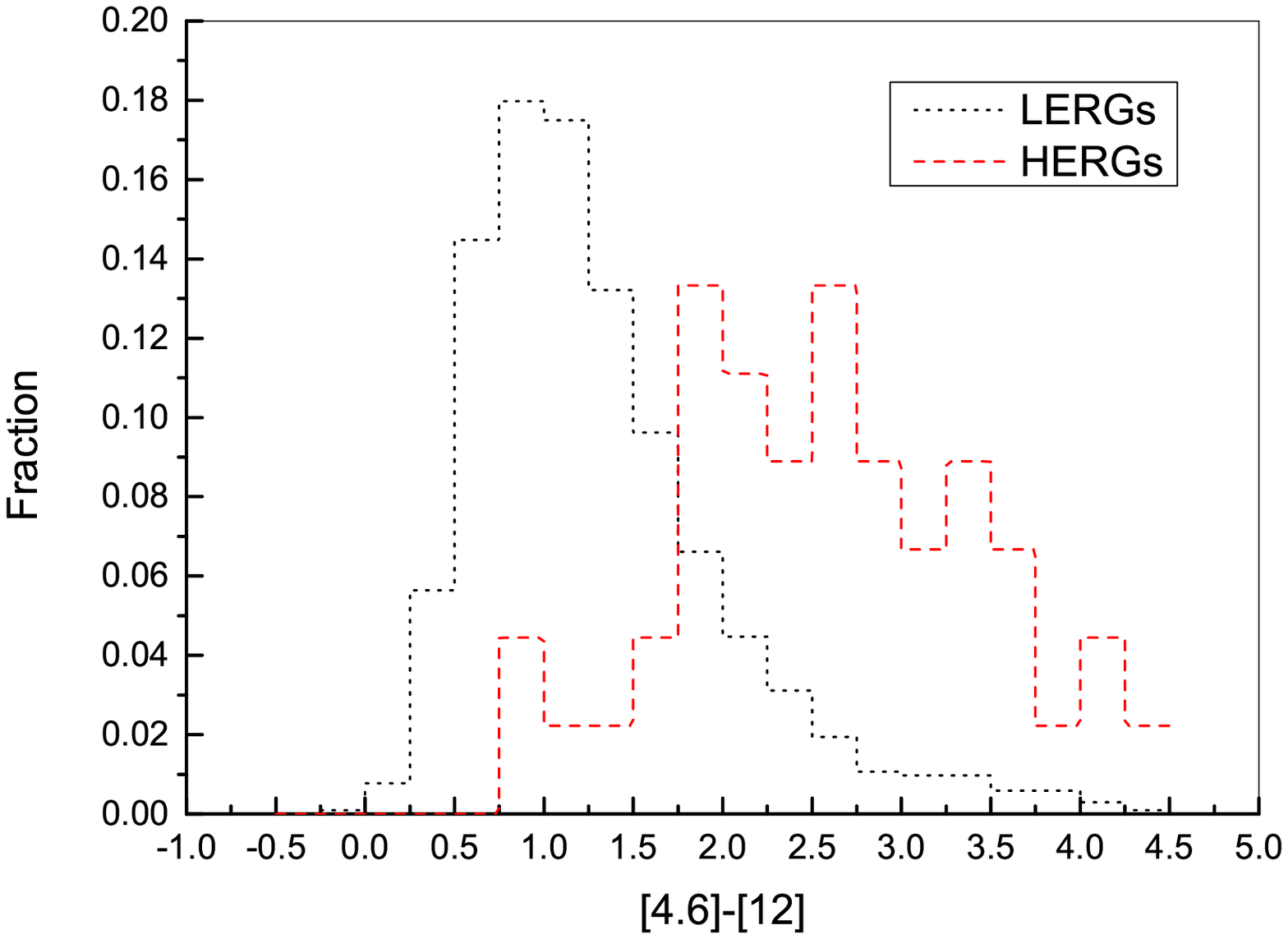}}}
\scalebox{0.5}[0.5]{\rotatebox{0}{\includegraphics[bb=50 40 531
436]{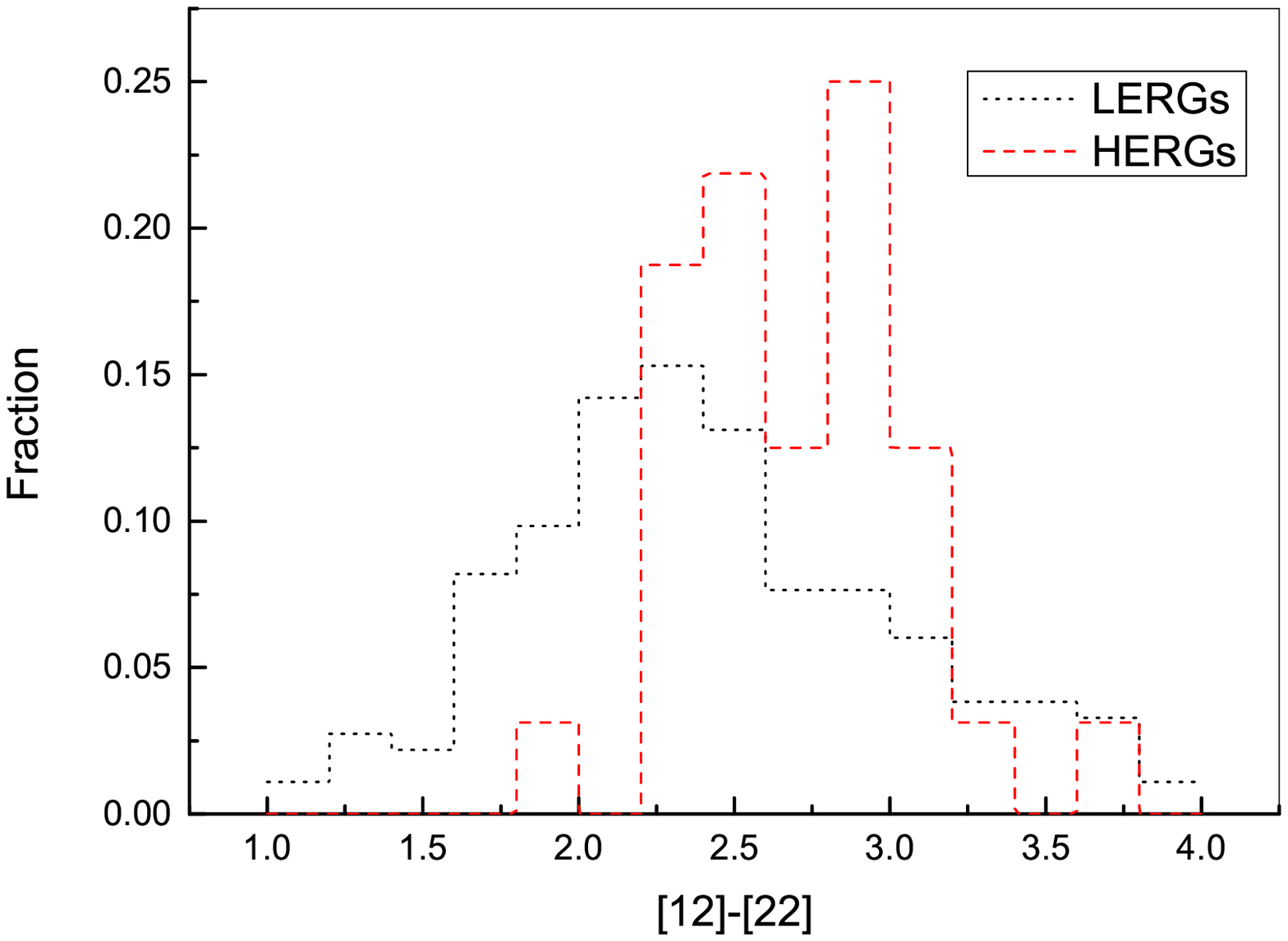}}}

\ \centering \caption{Normalized histograms of three colours of HERGs (red) and LERGs (black). Top panel is for colour $[3.4]-[4.5]$, middle panel is for colour $[4.6]-[12]$, and bottom panel is for colour $[12]-[22]$. The objects with upper limits are excluded.}
\label{fig 5}
\end{figure}

Chen \& Shan (2009) also reported that almost all type 1 quasars are distributed in the region around the PL line in the  $(J-H)-(H-K_s)$ diagram due to their predominantly non-thermal properties, whereas the majority of Sy2 galaxies are distributed in the region around the BB line due to the obscuring of the non-thermal component of AGNs by the dust torus. Here the  $(J-H)-(H-K_s)$ diagram shows that many Sy1 RLGs tend to lie close to the BB line. The difference in distribution of Sy1 RLGs and type 1 quasars occurs because that type 1 quasars have more luminous nuclei; therefore, the NIR emissions from the nuclei are significantly stronger than those from their host galaxies. It is noted that a great number of Sy2 RLGs are distributed in the region around the PL line. A similar case also exists for type 2 quasars (Chen \& Shan 2009). Chen \& Shan (2009) determined that about 43$\%$ of type 2 quasars are distributed in the region around the BB line, whereas about 57$\%$ of type 2 quasars are distributed near the PL line. They cited the possibility that some NIR emissions from the central nuclei can penetrate the dust torus because it is not substantially thick and the viewing angle is not very large. Another possible reason is that the torus is clumpy; therefore, the radiation from the central nuclei can penetrate the clumpy torus at a certain angle.

The middle and bottom panels of Figure 2 show the normalized histograms of the colours $J - H$ and $H-K_s$ of RL AGNs and SFGs, respectively. In addition, we performed a Kolmogorov-Smirnov (K-S) test to evaluate the similarity in the distributions of these colours for RL AGNs and SFGs. We determined that the similarity of $J-H$ has a significance level of less than 5$\%$, whereas that of $H-K_s$ is larger than 5$\%$. These results indicate that the distributions of $J-H$ for RL AGNs and SFGs are essentially different, whereas those of $H-K_s$ are essentially the same. Thus, a systematic difference exists in J-band emissions of RL AGNs and SFGs, the reason for which needs to be investigated.

\section{MIR colour properties}
\subsection{Colour properties}
Figure 3 shows $([3.4]-[4.6])-([4.6]-[12])$ and $([12]-[22])-([4.6]-[12])$ two-colour diagrams. In each diagram, a PL line and a BB line are plotted with a dashed and dotted line, respectively.

From Figure 3, we can see that RL AGNs and SFGs are clustered in distinctly separate regions. Compared to the RL AGNs, the SFGs have a redder $[4.6]-[12]$ colour and a slightly redder [3.4]-[4.6] colour. The PL spectral index $\alpha(3.4,4.6)$ of the RL AGNs changes from approximately 0 to 2, $\alpha(4.6,12)$ changes from approximately $-2$ to 2 and $\alpha(12,22)$ changes from approximately $-4$ to 0. This suggests that their observed spectra behave differently within the wavelength range of 4.6$-$22$\mu$m. One type is $\alpha(4.6,12)<0$ and $\alpha(12,22)<0$, showing an increase of $F_\lambda$ with $\lambda$ from 4.6 to 22 $\mu$m, and the other type is $\alpha(4.6,12)>0$ and $\alpha(12,22)<0$, showing a `turning point' in the spectra, i.e. $F_\lambda$ decreases with $\lambda$ from 3.4 to 12$\mu$m while $F_\lambda$ increases with $\lambda$ from 12 to 22$\mu$m. However, the SFGs show red spectra, i.e. $F_\lambda$ increases with $\lambda$ from 4.6 to 22$\mu$m.

To understand the MIR colours of the RLGs, a BB component and a PL component are used. The BB component roughly models the component of the stellar population, whereas the PL component represents the contributions of multi-temperature hot dusts or non-thermal emissions associated with the centre nuclei. The ultraviolet to MIR continuum of AGNs is described by a PL (Stern et al. 2005). The thermal emissions of multi-temperature hot dusts are also described by a PL, as seen in Figure 3 of Brandl et al. (2006). For example, Smith et al. (2007) used the modified blackbodies at fixed temperatures (35, 40, 50, 65, 90, 135, 200 and 300 K) as thermal dust continuum components to decompose the MIR spectra of SFGs. The two components are composed according to the fraction of their contribution to the $W$2-band flux, so we can calculate the flux densities of the composed model at different wavelengths by
\begin{equation}
F_{\lambda}=B_{\lambda}(T)+A_1\times \lambda^{-\alpha}.
\end{equation}
Here, $B_{\lambda}(\text{T})$ represents the Planck function of a BB with temperature $T$ and $A_1=\frac{\gamma_1}{1-\gamma_1} \frac{B_{\lambda_{4.6\mu m}}(T)}{\lambda_{4.6\mu m}^{-\alpha} }$, where $\gamma_1$ is the ratio of the radiation flux density of the PL component to the total flux density at the $W$2 band. For given values of temperature $T$ and spectral index $\alpha$, we can obtain the dependence of the colour of the composed model on $\gamma_1$ and plot the loci of the composed model with blue and red lines, as in Figure 4. Here, for simplicity, we assign the BB temperatures $T$ = 3500 and 5500 K, according to the later stellar effective temperature. We use different spectral indices $\alpha$ to model the potential energy distributions of emissions from multi-temperature dusts or non-thermal emissions associated with the centre nuclei. From the ([12]-[22])$-$([4.6]-[12]) diagram, we can see that the spectral indices $\alpha(12,22)$ of few sources are less than -4, so we set $\alpha \geq -4$, such as $\alpha = -2$, $-3$, and $-4$, respectively. Numbers along the loci are the fraction of the contribution of the PL component to the $W$2-band flux. Figure 4 also shows the probability distribution of LERGs and SFGs with grey-scale and 30 isodensity contour lines (black solid lines for LERGs and black dotted lines for SFGs). The probability distribution is evaluated using the kernel density estimation technique, which is an effective method of estimating the probability function of a multivariate variable without any assumption. For comparison, we also superimpose HERGs as red circles in Figure 4. From Figure 4, we can see that: the smaller the spectral index $\alpha$, the redder the colour [4.6]-[12]; the larger the contribution of the PL component to the W2-band flux, the closer to the PL line the distribution of the RLGs in the ([4.6]-[12])-([12]-[22]) diagram is. Although our parameters cannot cover the distribution of all the RLGs, the composed model helps us to understand the role of the BB and PL components in the MIR colours of RLGs.

Figure 4 also shows that LERGs have a distribution of a double-core structure, which means that LERGs do not have singular infrared properties. The main difference between the two cores is their different $[3.4]-[4.6]$ colour. More spectral data are required to understand this difference. It is noted that the $W$1 band covers the PAH emission at 3.3 $\mu$m and the $W$3 band covers the 9.8-$\mu$m silicate absorption and the 11.3-$\mu$m PAH emission, while there are no PAH emissions in the $W$2 and $W$4 bands (e.g. Brandl et al. 2006; Smith et al. 2007). Therefore, many LERGs display a bluer $[3.4]-[4.6]$ colour than the red line in Figure 4, due to the 3.3-$\mu$m PAH emission.

\begin{figure}

\scalebox{0.5}[0.5]{\rotatebox{0}{\includegraphics[bb=50 20 500
360]{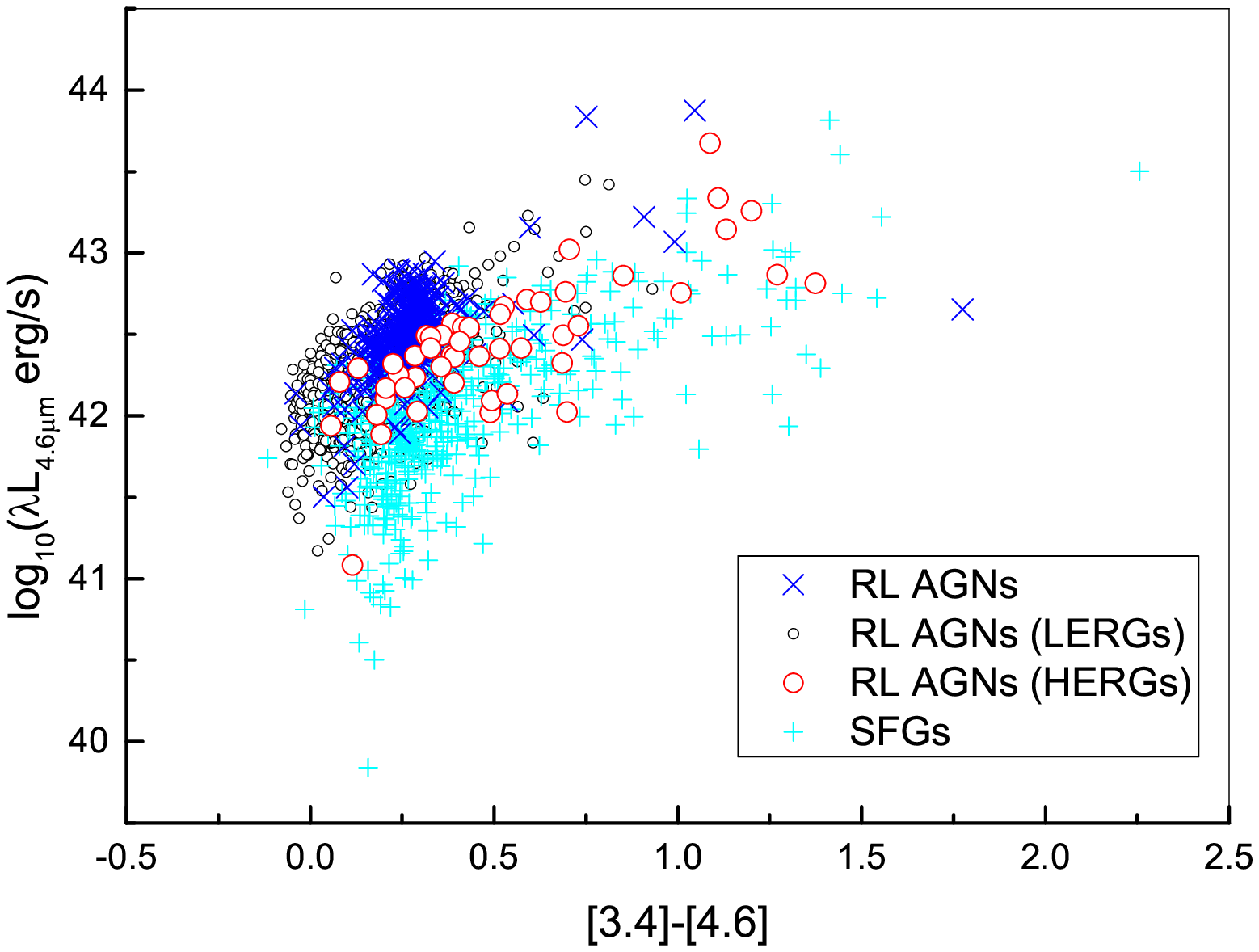}}}
\scalebox{0.5}[0.5]{\rotatebox{0}{\includegraphics[bb=50 20 500
360]{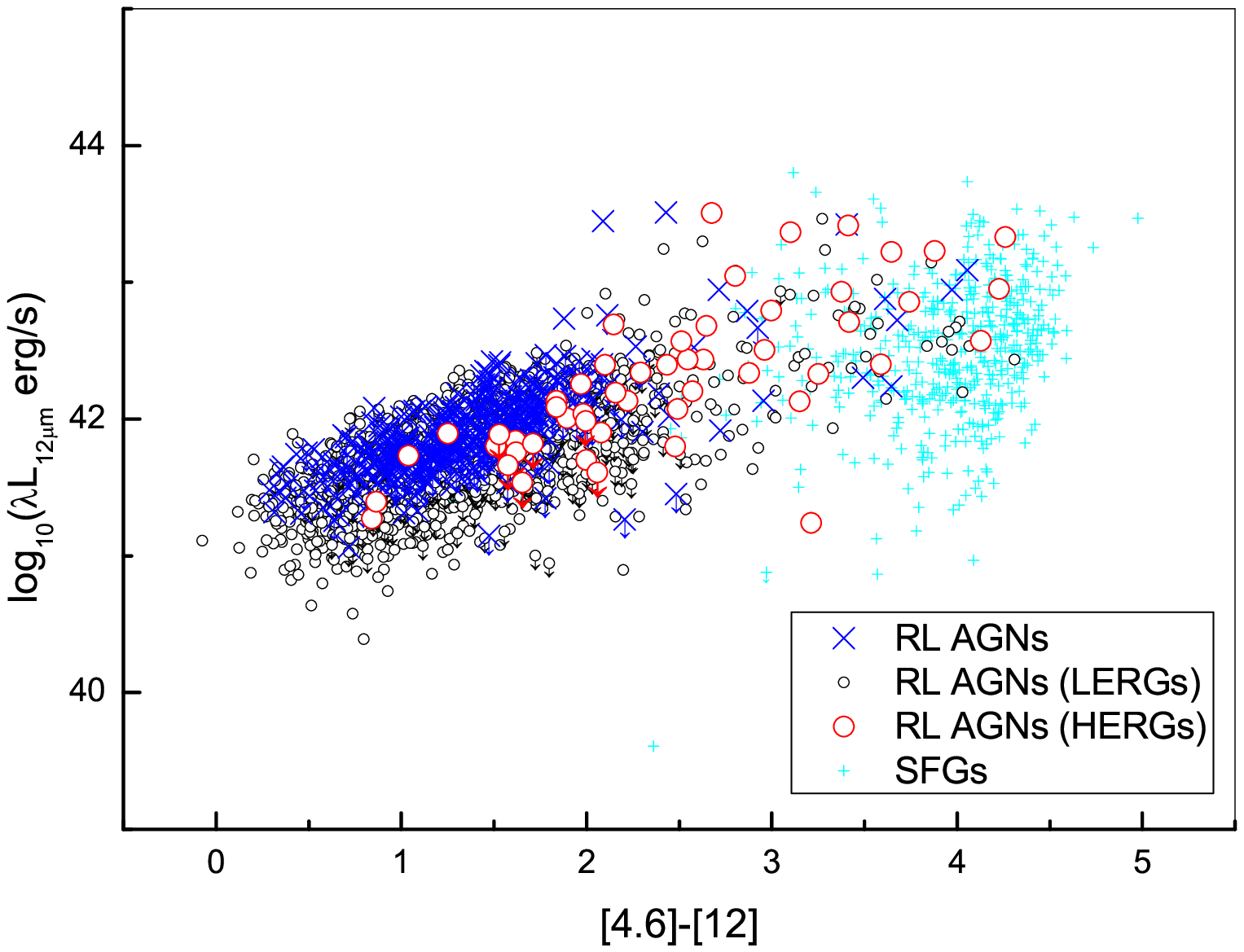}}}
\scalebox{0.5}[0.5]{\rotatebox{0}{\includegraphics[bb=50 20 500
360]{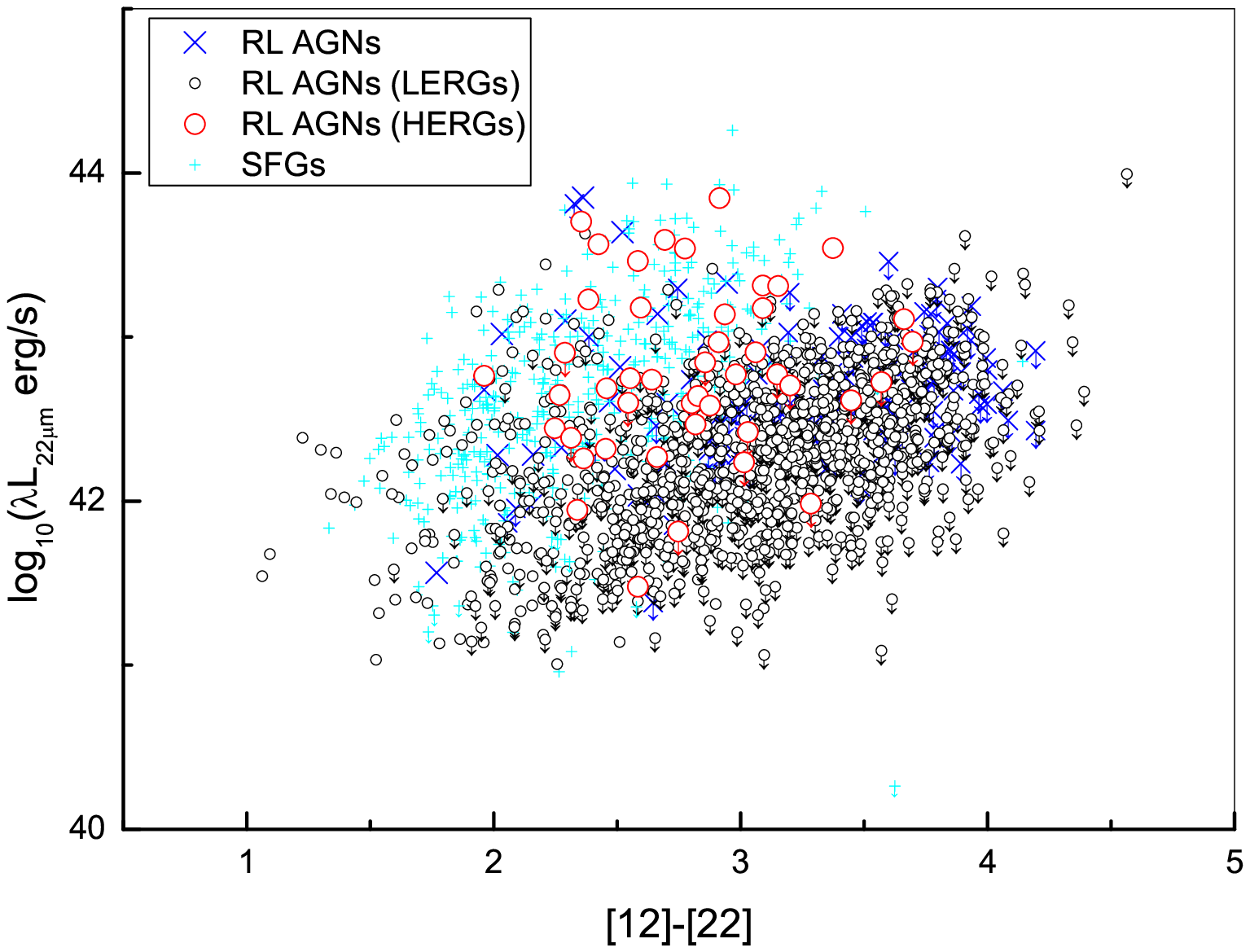}}}

\ \centering \caption{Plots of the logarithms of 4.6-$\mu$m luminosity versus the colour $[3.4]-[4.6]$ (top panel), the logarithms of 12-$\mu$m luminosity versus the colour $[4.6]-[12]$ (middle panel), and the logarithms of 22-$\mu$m luminosity versus the colour $[12]-[22]$ (middle panel).}
\label{fig 6}
\end{figure}

Similarly composed models are used to mimic the distribution of most RL AGNs and SFGs in the $([12]-[22])-([22]-[12])$ diagram. Although the $([3.4]-[4.6])-([4.6]-[12])$ and $([12]-[22])-([22]-[12])$ diagrams cannot definitively confirm the nature of thermal or non-thermal emissions of RL AGNs and SFGs at MIR wavelengths, they help us understand the spectral shapes of RL AGNs and SFGs. For SFGs, we presume that multi-temperature hot dusts dominate MIR emissions, as what Brandl et al. (2006) pointed out. The $([12]-[22])-([22]-[12])$ diagram also shows that the $[12]-[22]$ colour of RL AGNs is more widely distributed than that of the SFGs, which suggests that the MIR radiation components of RL AGNs is more diverse than that of SFGs. MIR radiation of RL AGNs may include the non-thermal emissions associated with AGNs and the thermal emissions from torus or host galaxies. The two responsible factors need be further quantitatively studied for RL AGNs. Stern et al. (2012) and Assef et al. (2013) presented a simple MIR colour criterion of selecting AGNs $[3.4]-[4.6]$ $>$ 0.8. In our studied samples, 0.06$\%$ of RL ANGs (14 RL AGNs) and 10$\%$ of SFGs (50 SFGs) achieve this criterion.

In addition, from the $([3.4]-[4.6])-([4.6]-[12])$ and $([12]-[22])-([4.6]-[12])$ diagrams, we can see the obvious difference in the distribution of LERGs and HERGs, though LERGs and HERGs partly overlap. To show the difference, we plot normalized histograms of three MIR colours ($[3.4]-[4.6]$, $[4.6]-[12]$, and $[12]-[22]$) of LERGs and HERGs in Figure 5. This plot shows that HERGs have redder $[3.4]-[4.6]$ and $[4.6]-[12]$ colours on average. LERGs are mainly located in the region of $[3.4]-[4.6]$ $<$ 0.4 and $[4.6]-[12]$ $<$ 0.8, whereas a large number of HERGs have $[3.4]-[4.6]$ $>$ 0.4 and $[4.6]-[12]$ $>$ 0.8. In particular, few HERGs have $[3.4]-[4.6]$ $<$ 0.1 and $[4.6]-[12]$ $<$ 0.8. HERGs have a narrower colour [12]-[22] distribution than LERGs. Figures 3 and 4 show that HERGs are distributed in the region where MIR emissions are dominated by the power-law component. Figures 4 also shows that a large number of LERGs are distributed in the region dominated by the steller population and a small number of LERGs have redder $[4.6]-[12]$ colour. The difference between LERGs and HERGs may be concerned with their different accretion modes. It is suggested that HERGs and LERGs have different accretion modes (Hardcastle et al. 2007). LERGs are suggested to be fuelled via ADAFs (hot-mode accretion) with low accretion rates, whereas HERGs are associated with a standard thin accretion disc (cold-mode accretion) at high accretion rates (Shakura \& Sunyaev 1973; Hardcastle et al. 2007; Best \& Heckman 2012). HERGs with high accretion rate have luminous core, while LERGs with low accretion rate have dim core. Therefore, the MIR emissions of HERGs are dominated by non-thermal emissions associated with AGNs; the MIR emissions from central core of LERGs are significantly contaminated by stars and dusts of host galaxies. Accretion rates play an important role in the switch between hot- and cold-mode accretions (Esin et al. 1997). However, accretion rates are also modulated by feedback from central AGNs, and different accretion modes of the central AGNs have different effects on this feedback (Ciotti \& Ostriker 2001; Gan et al. 2014). Determining what the key of switch is between hot- and cold-mode accretions is still an open issue.


\begin{table}
\begin{center}

\caption[]{Results of partial correlation analyses between luminosities and colours.} \label{table2}\tiny
\begin{tabular}{cccccc}

\hline\noalign{\smallskip}
 Abscissa & Ordinate       & Subsample & Number & $\rho_{.z}$ & $\tau/\sigma$ \\
\hline\noalign{\smallskip}
$[3.4]-[4.6]$ & $\text{log}_{10}(\lambda L_{4.6\mu m})$ &  LERGs & 1748 & 0.459 & 8.80 \\
            &                                         & HERGs & 52   & 0.585 & 6.50 \\
            &                                         & SFGs  & 502  & 0.460 &11.94 \\
                                                        \hline\noalign{\smallskip}
$[4.6]-[12]$ & $\text{log}_{10}(\lambda L_{12\mu m})$ & LERGs & 1748 & 0.284 & 27.80 \\
           &                                        & HERGs & 52   & 0.617 & 13.0 \\
           &                                        & SFGs  & 502  & 0.197 &11.39 \\
                                                        \hline\noalign{\smallskip}
$[12]-[22]$ & $\text{log}_{10}(\lambda L_{22\mu m})$ & LERGs & 1748 & 0.027 & 7.90 \\
          &                                        & HERGs & 52   & 0.179 & 3.35\\
          &                                        & SFGs  & 502  & 0.388 &17.29 \\

\hline\noalign{\smallskip}
\end{tabular}

\begin{list}{}
\item\scriptsize{Notes. $\rho_{.z}$ is Kendall's partial correlation coefficient with its associated null probability. $\tau/\sigma>3$ means that the correlation is significant at the $3\sigma$ level.}
\end{list}

\end{center}
\end{table}
\subsection{Luminosity versus colour}

Here, we check possible correlations between MIR luminosities and colours for different types of RLGs. We calculate the luminosity distance ($D_{\text{L}}$) by using the redshift listed in Table 1 (cosmology parameters are assumed in Section 1) and then obtain the absolute luminosities. The $\textit{K}$ correction may become important for high-redshift objects. We apply the \textit{K} correction to our sample. For simplicity, MIR spectra of our sample are approximated by simple PL form ($\propto\nu^{\alpha}$). The intrinsic luminosities $L(\nu)$ are calculated as follows: $L(\nu)=4\pi D_{\text{L}}^2 f(\nu) (1+z)^{-\alpha-1}$ (e.g. Bourne et al. 2011 ), where $f(\nu)$ is the observed flux density at the observing frequency $\nu$. Here, $\alpha$ can be calculated by using Equation 1. In Figure 6, we plot the logarithms of 4.6-$\mu$m, 12-$\mu$m, and 22-$\mu$m luminosities versus the $[3.4]-[4.6]$, $[4.6]-[12]$, and $[12]-[22]$ colours, respectively. We can see that the 12-$\mu$m and 22-$\mu$m luminosities of HERGs are on average higher than those of LERGs. Similar results are also shown in Figure 7 of G\"{u}rkan et al. (2014). G\"{u}rkan et al. (2014) remarked that a division of LERGs and HERGs clearly stands out in the 22-$\mu$m luminosity. However, based on our studied sample, HERGs cannot be separated from LERGs using the 22-$\mu$m luminosity.

In order to test physical correlations between luminosities and
colours, we further calculate Kendall's partial correlation
coefficient ($\rho ._{z}$) with censored data using the methods by
Akritas \& Siebert (1996). $\rho ._{z}$ measures the degree of
association between two random variables, with the effect of a set
of controlling random variables removed. $\rho ._{z}$  is especially
important for excluding the effect of redshift. The values of $\rho
._{z}$ are within $-1$ and 1. The larger their absolute values, the
stronger the correlations. The positive values indicate the positive
correlation and the negative values indicate the negative
correlation. Table 3 gives results of partial correlation analyses.
For LERGs, a physical relationship presents between
$\text{log}_{10}(\lambda L_{4.6\mu m})$ and $[3.4]-[4.6]$ and the
correlation of luminosities vs. colours becomes weak with increasing
wavelength. As shown in the bottom panel of Figure 6, no correlation
between $\text{log}_{10}(\lambda L_{22\mu m})$ versus $[12]-[22]$ is seen
for LERGs. For HERGs, good physical correlations exist between
$\text{log}_{10}(\lambda L_{4.6\mu m})$ versus $[3.4]-[4.6]$ and between
$\text{log}_{10}(\lambda L_{12\mu m})$ versus $[4.6]-[12]$, but the
correlation of  $\text{log}_{10}(\lambda L_{22\mu m})$ versus $[12]-[22]$
is weaker. For SFGs, weak correlations present between
$\text{log}_{10}(\lambda L_{4.6\mu m})$ versus $[3.4]-[4.6]$ and between
$\text{log}_{10}(\lambda L_{22\mu m})$ versus $[12]-[22]$.

\begin{figure}
\scalebox{0.5}[0.5]{\rotatebox{0}{\includegraphics[bb=50 20 500
360]{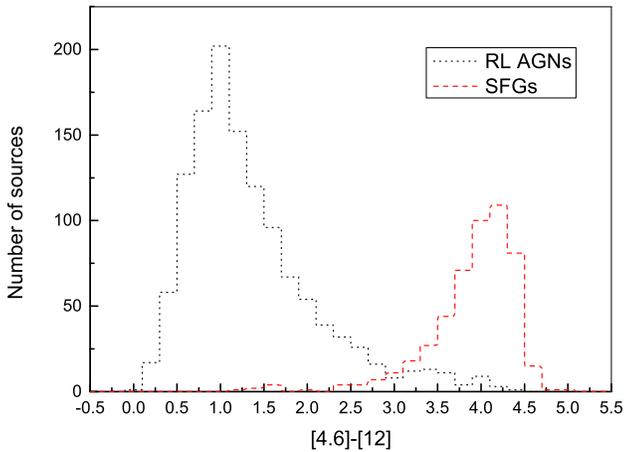}}}
 \ \centering \caption{Histogram of colour $[4.6]-[12]$ of RL AGNs (black dotted line) and SFGs (red dashed line). The objects with upper limits are excluded.}
\label{fig 7}
\end{figure}

\subsection{Distinguishing RL AGNs and SFGs by MIR colours}

The activity in galaxies is thought to be influenced by two components, star formation and AGNs. Some galaxies are dominated by either star formation or the AGN, while others exhibit comparable contributions from both components. The AGN of NGC 1068 is surrounded by a ring of star formation that contributes half of the total luminosity of NGC 1068 (Telesco \& Decher, 1988). The two components are important in the understanding of galaxy energetics and galaxy evolution. It is necessary to distinguish the dominant component of the activity for active galaxies. Both starburst and Sy2 AGN galaxies exhibit emission lines with FWHM$<$ 1000 km s$^{-1}$. However, Baldwin et al. (1981), Veilleux \& Osterbrock (1987) and Kauffmann et al. (2003) demonstrated that Sy2 AGN galaxies may be distinguished from normal SFGs based on the diagnostic diagram of [O III]/H$\beta$ versus [HII]/H$\alpha$, though not all galaxies can be easily segregated. Machalski \& Condon (1999) separated the RL AGN and star-forming populations on the basis of FIR-radio flux ratios, FIR spectral indices and radio-optical flux ratios. Best et al. (2005) pointed out that RL AGNs are fairly well separated from SFGs in the plane of $D_{n}(4000)$ versus
$L_{1.4GHz}/M_{*}$.

Based on the classification provided by Best et al. (2005) and Best \& Heckman (2012), we plot a histogram of the $[4.6]-[12]$ colour in Figure 7. Figure 7 shows the separation of RL AGNs from SFGs; 95$\%$ of RL AGNs have $[4.6]-[12]$ $<$ 3.0, and 94$\%$ of SFGs have $[4.6]-[12]$ $>$ 3.0. The colours $[3.4]-[12]$, $[3.4]-[22]$, and $[4.6]-[22]$ have similar functions in the separation of the two types of galaxies, which is attributed to the fact that the 12-$\mu$m and 22-$\mu$m emissions of SFGs are redder than those of RL AGNs.

\section{FIR colour properties}
We search \textit{AKARI}-FIS PSC and obtain \textit{AKARI}-FIS data with high quality. \textit{AKARI}-FIS data have four-level flux quality indicator. High quality, i.e. the flux quality indicator equals 3, means that the source is confirmed and the flux is reliable. We neglect the data with the flux quality indicator equal to 1 and 2, which means that the source is not confirmed or the flux is not reliable. 15 and 282 \textit{AKARI} 90-$\mu$m counterparts are obtained for RL AGNs and SFGs, respectively. The detection rate of RL AGNs at the 90-$\mu$m band is $\sim 0.7\%$, which is much less than $\sim 56 \%$ of SFGs. The majority of the RL AGNs studied here are not included in the \textit{AKARI}-FIS PSC, though the \textit{AKARI}-FIS survey covered more than 94$\%$ of the entire sky, and the \textit{AKARI}-FIS PSC includes 0.4 million infrared sources brighter than the detection limit (approximately 0.6 Jy) at 90 $\mu$m (Yamamura et al. 2010). This suggests that the 90 $\mu$m flux density of most of RL AGNs is lower than 0.6 Jy. In addition, 15 \textit{AKARI} 90-$\mu$m counterparts of RL AGNs include nine LERGs and one HERG, respectively. The detection rate of LERGs and HERGs reach $\sim$0.5 and $\sim$2 per cent at 90-$\mu$m band, respectively. The FIR emissions at 90 and 140 $\mu$m are thought be associated with cold dust in host galaxies (e.g. Netzer et al. 2007; Mullaney et al. 2011), and then HERGs tend to have more abundant cold dust than LERGs, on average.

\begin{figure}

\scalebox{0.5}[0.5]{\rotatebox{0}{\includegraphics[bb=55 23 500
360]{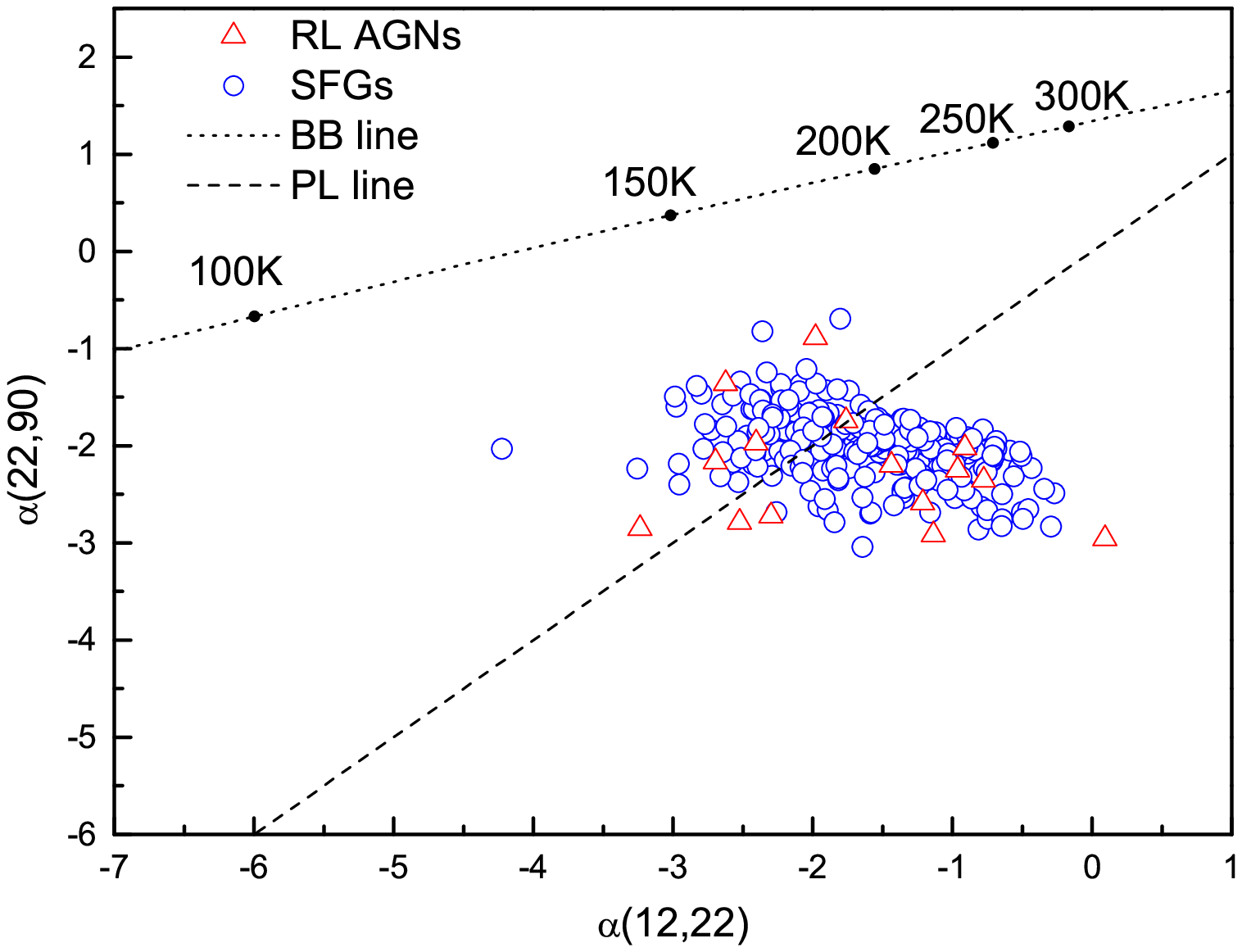}}}

\scalebox{0.5}[0.5]{\rotatebox{0}{\includegraphics[bb=55 23 500
380]{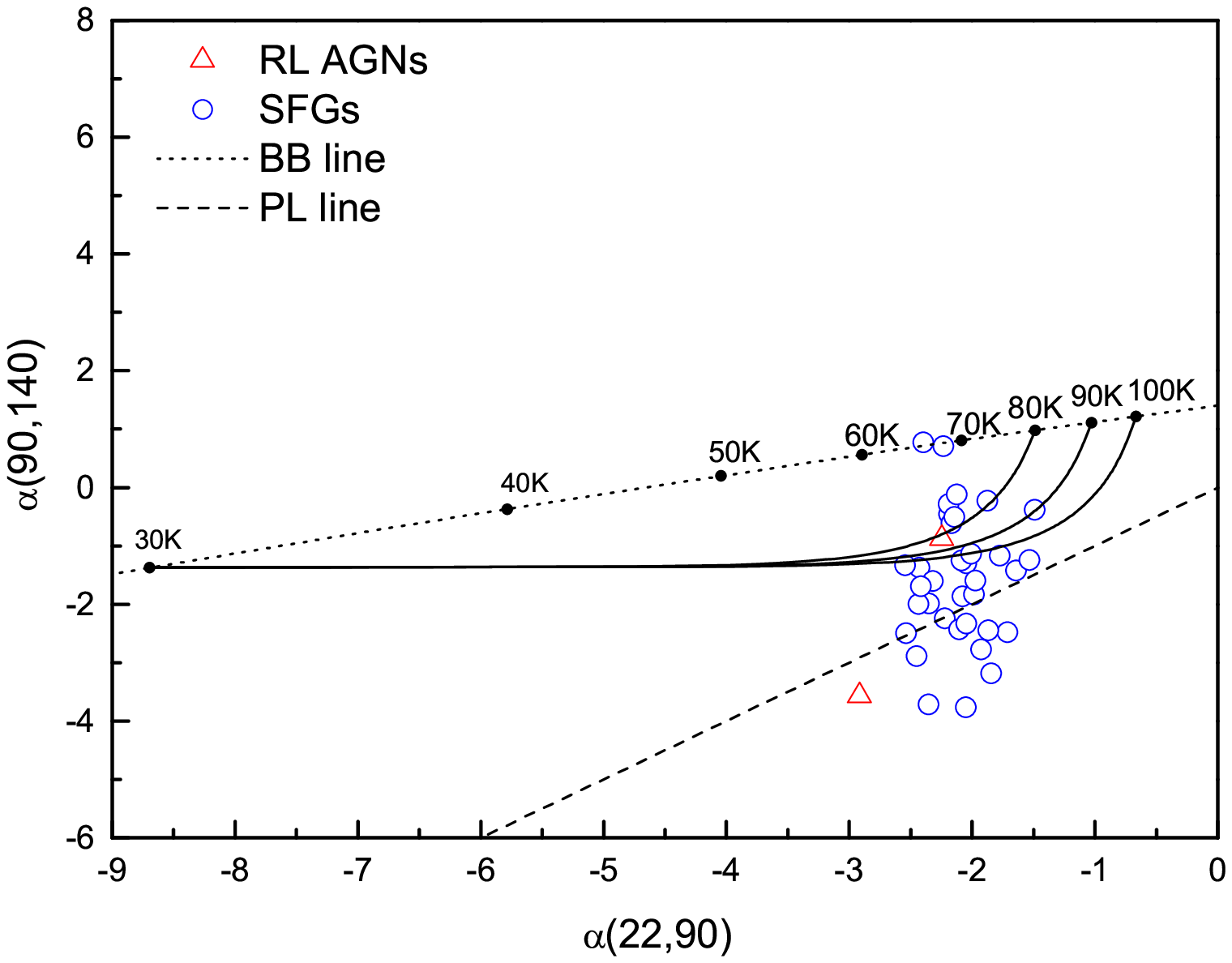}}}

\ \centering \caption{The $\alpha(22,90)-\alpha(12,22)$ and $\alpha(90,140)-\alpha(22,90)$ diagrams. A BB line marked with temperature and a PL line are also plotted as the dashed and dotted line, respectively. A set of solid lines in the $\alpha(90,140)-\alpha(22,90)$ diagrams indicate the loci of models with a cold dust component of 30 K and a warm dust component of 80$-$100 K. Circles represent RL AGNs and triangles represent SFGs.}
\label{fig 8}
\end{figure}

The $\alpha(22,90)-\alpha(12,22)$ and $\alpha(90,140)-\alpha(22,90)$ diagrams are also shown in Figure 8 with PL and BB lines plotted as dashed and dotted lines, respectively. A set of solid lines in the $\alpha(90,140)-\alpha(22,90)$ diagram indicate the loci of the composed model, with a cold disc component of $\sim$30 K and a warm dust component of 80$-$100 K. The two components are composed according to the fractional contribution of the warm dust component to the 90 $\mu$m-band flux. We can calculated the flux densities at a given wavelength by
\begin{equation}
F_{\lambda}=B_{\lambda}(T_\text{c})+A_2\times B_{\lambda}(T_\text{w}).
\end{equation}
Here, $B_{\lambda}(T_\text{c})$ and $B_{\lambda}(T_\text{w})$ represent the Planck functions of the cold component with temperature $T_\text{c}$ and the warm component with $T_\text{w}$, respectively, and $A_2=\frac{\gamma_2 B_{90\mu m}(T_{\text{c}})}{(1-\gamma_2) B_{90\mu m} (T_{\text{w}})}$, where $\gamma_2$ is the ratio of the radiation flux density of the warm component to the total flux density at 90 $\mu$m. We can then draw the set of black line in the $\alpha(90,140)-\alpha(22,90)$ diagram by computing the spectral index between two bands from Equation 1 when $\gamma_2$ changes.

Figure 8 shows $\alpha(22,90)<0$ for all RL AGNs detected at 90$\mu$m, which indicates that some RL AGNs have a red FIR colour. In combination with Figure 3, we can see that the spectral indices for the majority of SFGs are $\alpha(3.4,4.6)>$0, $\alpha(4.6,12)<$0, $\alpha(12,22)<$0, $\alpha(22,90)<$0 , and $\alpha(90, 140)<$0, i.e. their spectral indices do not cross zero, which suggests that SFGs have similar spectra from 3.4 to 90$\mu$m. The spectra may be described by a PL with a similar spectral index, as shown by the 5$-$38$\mu$m MIR spectra of 22 starburst nuclei presented by Brandl et al. (2006, see Figure 3 in their paper). However, the bottom panel in Figure 8, the $\alpha(90,140)-\alpha(22,90)$ diagram, shows that a small number of SFGs with $\alpha(90,140)<-1.4$ are distributed below the black solid lines and close to the PL line. A set of black solid lines, plotted in the $\alpha(90,140)-\alpha(22,90)$ diagram, indicates the composed model of a cold disc component ($\sim$30 K) and a warm dust component. As shown in the $\alpha(90,140)-\alpha(22,90)$ diagram, the sources above the black solid lines and close to the BB line may be explained by thermal emissions from dust at MIR wavelengths, whereas the MIR properties of the source distributed in the region around the PL line cannot be explained by thermal emissions of the composed model, meaning that non-thermal emissions may be used to understand their MIR properties. In general, the FIR emissions of SFGs are thought to be caused by the heating of dust by young, massive stars (Condon et al. 1991; Helou \& Bicay 1993; Machalski \& Condon 1999). The $\alpha(90,140)-\alpha(22,90)$ diagram suggests that there are many SFGs whose FIR emissions may not come from dust. We also suggest the criteria $\alpha(90,140)<-1.4$ to select active galaxies with non-thermal emissions at MIR wavelengths.



\section{Summary}

The radio emissions of RLGs are dominated either by AGNs or by supernova remnants in their star-forming regions; however, both the dusty tori associated with the AGNs and the dust and gas clouds heated by young stars emit MIR or FIR emissions. Here, we analyse the infrared colour properties of a large sample of RLGs based on the 2MASS, WISE and AKARI survey data. RL AGNs and SFGs are separately distributed in a $([3.4]-[4.6])-([4.6]-[12])$ diagram, which provides the colour $[4.6]-[12]$ to separate RL AGNs from SFGs. 95$\%$ of RL AGNs have $[4.6]-[12]$$<$ 3.0, and 94$\%$ of SFGs have $[4.6]-[12]$$>$3.0. This difference may be due to the significant contribution of the dusts of SFGs to MIR emission. It is noted that a small number of RL AGNs are distributed in the region of SFGs in the $([3.4]-[4.6])-([4.6]-[12])$ diagram, which indicates that the RL AGNs in the region of SFGs have redder MIR and FIR colours than ordinary RL AGNs. The 3.4-90$\mu$m observed spectra of SFGs are described by similar spectra, like a PL, which are indicated by the distribution of SFGs around the PL line in $([12]-[22])-([4.6]-[12])$ and $\alpha(90,140)-\alpha(22,90)$. However, the 3.4$-$22$\mu$m spectra of RL AGNs shows different types, due to the variation of $\alpha(4.6,12)$ from $-2$ to 2. One type has the flux density increasing with wavelength from 3.4 to 22$\mu$m; the other type has the flux density decreasing with wavelength from 3.4 to 12$\mu$m, whereas flux density increases with wavelength from 12 to 22$\mu$m.

The MIR colours of RLGs are described by a model with a BB component and a PL component. However, the PL component cannot exclude the contribution from thermal emissions, because the PL component may be described by multi-colour BB spectra, indicating dusts with various temperatures.

LERGs and HERGs have different distributions of the $[3.4]-[4.6]$, $[4.6]-[12]$ and $[12]-[22]$ colours, which means that LERGs and HERGs have intrinsically different MIR properties.

The $\alpha(90,140)-\alpha(22,90)$ diagram indicates that many SFGs with $\alpha(90,140)<-1.4$ are distributed in the region around the PL line. They cannot be described by the composed model with a cold disc component ($\sim$30 K) and a warm dust component involved in star formation. The FIR emissions of the SFGs with $\alpha(90,140)<-1.4$ may not originate from dusts but instead from non-thermal processes. Therefore, the criterion $\alpha(90,140)<-1.4$ provides a way to select active galaxies, for which non-thermal emissions may contribute to FIR wavelength.

\section*{Acknowledgements}
We thank the anonymous referee for useful comments to improve this paper.
We also thank J.H. He for valuable discussions.
This work was supported by the Fundamental Research Funds for the
Central Universities (no.CQDXWL-2014-004). Chen, P.S. is supported
by the Natural Science Foundation of China (grants 11173056).

This publication makes use of data products from the 2MASS,
which is a joint project of the University of Massachusetts and the Infrared
Processing and Analysis Center/California Institute of Technology,
funded by the National Aeronautics and Space Administration and the National Science Foundation,
and from the WISE,
which is a joint project of the University of California, Los Angeles,
and the Jet Propulsion Laboratory/California Institute of Technology,
funded by the National Aeronautics and Space Administration.
This research is based on observations with \textit{AKARI}, a JAXA project with the participation of ESA.


\clearpage

\clearpage

\end{document}